\documentclass[A4paper]{article}  
\usepackage{moreverb,url}

\usepackage[colorlinks,bookmarksopen,bookmarksnumbered,citecolor=black,urlcolor=red]{hyperref}
\usepackage[utf8]{inputenc}

\usepackage{graphicx}

\usepackage[margin=1in]{geometry} 

\usepackage{amsmath}
\usepackage{amsthm}
\usepackage{amssymb}

\newcommand\BibTeX{{\rmfamily B\kern-.05em \textsc{i\kern-.025em b}\kern-.08em
T\kern-.1667em\lower.7ex\hbox{E}\kern-.125emX}}

\begin{document}

\title{Online control of the False Discovery Rate in group-sequential platform trials}

\author{Sonja Zehetmayer\footnote{sonja.zehetmayer@meduniwien.ac.at}, Martin Posch, Franz Koenig}

\date{Center for Medical Statistics, Informatics and Intelligent Systems, Medical University of Vienna, Austria
}

\maketitle

\begin{abstract}
When testing multiple hypotheses, a suitable error rate should be controlled even in exploratory trials.
Conventional methods to control the False Discovery Rate (FDR) assume that all p-values are available at the time point of test decision. In platform trials, however, treatment arms enter and leave the trial at any time during its conduct. Therefore, the number of treatments and hypothesis tests is not fixed in advance and hypotheses are not tested at once, but sequentially. Recently, for such a setting the concept of online control of the FDR was introduced.

We investigate the LOND procedure to control the online FDR in platform trials and propose an extension to allow for interim analyses with the option of early stopping for efficacy or futility for individual hypotheses. 
The power depends sensitively on the prior distribution of effect sizes, e.g., whether true alternatives are uniformly distributed over time or not. We consider the choice of design parameters for the LOND procedure to maximize the overall power and compare the O'Brien-Fleming group-sequential design with the Pocock approach. Finally we investigate the impact on error rates by including both concurrent and non-concurrent control data.   \end{abstract}

\maketitle

\section{Introduction}
Platform trials are an innovative type of study design, where randomized clinical trials with related aims or questions are combined to improve efficiency by reducing costs or saving time \cite{meyer2020evolution,berry2015platform}. Treatment arms can enter and leave the trial at any time during its conduct and the total number of hypothesis tests is not fixed in advance. One major benefit of platform trials is the comparison of treatment arms to one shared control arm and a therefore reduced number of control patients. As more than one statistical test is performed, an adjustment for testing multiple hypotheses has been proposed \cite{Wason21,Bai20}, e.g., control of the Family Wise Error Rate (FWER) or the False Discovery Rate (FDR), defined as the expected proportion of false rejections under all rejections \cite{Benjamini95}. Conventional methods, however, as the Bonferroni method to control the FWER or the Benjamini-Hochberg (BH) method to control the FDR, require that the number of hypothesis tests is fixed and, for the BH method, that all p-values are available at the time point of test decision. These methods are thus not appropriate for the special design of platform trials. Recently, the concept of online control of the FDR \cite{Javanmard15,Javanmard18} or the FWER \cite{Ramdas21} was introduced  where hypothesis tests and test decisions can be performed sequentially while guaranteeing on FDR control. At each step a decision has to be performed if the current null hypothesis should be rejected based on previous decisions but without knowledge on future p-values or the number of hypotheses to be tested.

To control the online FDR, several procedures have been proposed for independent test statistic, as e.g., the LORD  \cite{Javanmard18}, or the SAFFRON method \cite{Ramdas18}. For these, however, it is so far not proven that they control the online FDR for positively dependent test statistics as it is the case in platform trials due to a shared control group. For the LOND procedure (significance Levels based On Number of Discoveries \cite{Javanmard15}) a proof has been given that the FDR is controlled also for positively dependent test statistics\cite{Zrnic21}, throughout the paper online control of the FDR will thus be performed with the LOND procedure. A detailed comparison of online FDR procedures can be found, e.g, in Robertson and Wason (2018)\cite{Robertson18} or Robertson et al. (2021) \cite{Robertson21} (see also the R-package onlineFDR \cite{onlinefdr}).

The aim of this paper is to explore the LOND procedure for platform trials and to present a framework for incorporating group-sequential hypothesis tests with the option of early efficacy or futility stopping for individual hypotheses while controlling the online FDR. In Section 2 we review the LOND procedure and propose an extension for the online FDR control of group-sequential designs (gsLOND) as well as two modifications for gsLOND. The procedures are investigated in simulation studies in Section 3. The average power, defined as the proportion of rejected alternatives among all alternatives will be analysed for different priors for the distribution of the alternative hypotheses or several methods distributing the alpha level among the hypotheses of interest. Scenarios including both concurrent and non-concurrent control data will be investigated. 

\section{Methods}
Consider a sequence of null hypotheses $H_1,H_2,\dots$ and corresponding p-values $p_1,p_2,\dots$, where the maximum number is not pre-fixed in advance. 
At each step $i$, a decision has to be performed whether to retain or reject $H_i$ with only having information on the previous hypotheses $H_1,\dots,H_{i-1}$ and without information on the eventually total number of hypotheses to be tested in the platform. Thus, in online testing, the significance level $\alpha_i$ for hypothesis $i$ is only a function of the decisions for $H_1,\dots,H_{i-1}$. Online FDR algorithms aim at FDR control at a pre-specified  significance level $\alpha$ after each test decision \cite{Zrnic21}.

\subsection{The LOND procedure - Significance levels based on number of discoveries}

Given an overall significance level $\alpha$, a sequence of non-negative numbers $\beta_i$ 
is determined before starting the trial, such that $\sum_{i=1}^\infty \beta_i=\alpha$. 

In the LOND procedure, the values of the nominal significance levels $\alpha_i$ for $H_i$ are given by:
\begin{equation}
\alpha_i=\beta_i(\sum_{j=1}^{i-1} R_j+1) \label{LOND}
\end{equation}

with test decision $R_j=1$ if $p_j\leq \alpha_j$ (reject $H_j$) and 0 otherwise.

It has been proven\cite{Zrnic21}, that this procedure controls the FDR after every decision step for independent or positively dependent p-values. 

\begin{figure}[h!]
\centering
\includegraphics[width=.7\textwidth]{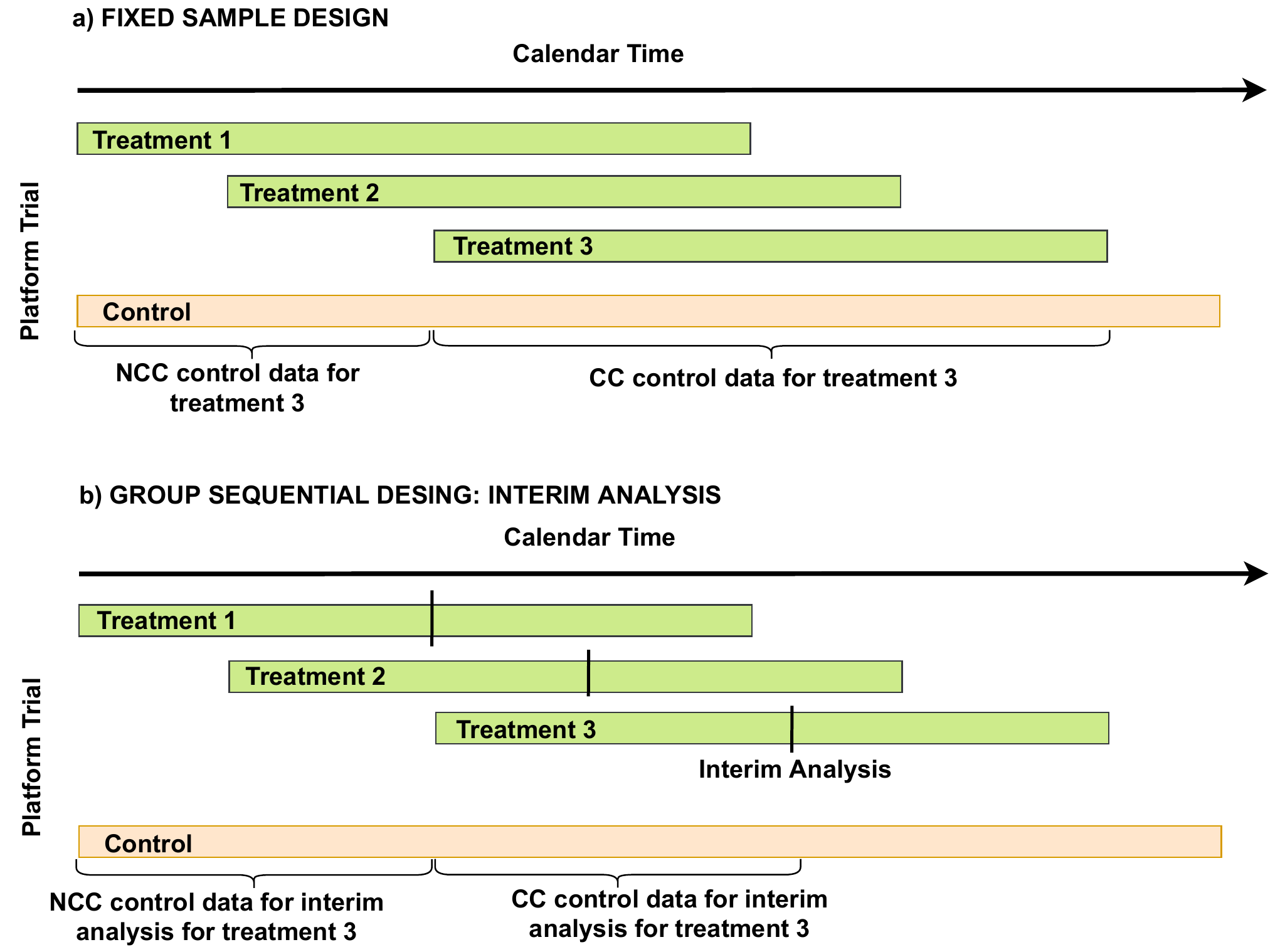}
\caption{a) Example of a platform trial with 3 treatments and one common control (fixed sample design) b) Example of a platform trial with a group-sequential design (illustration of one interim analysis for each treatment). Two strategies are possible: Either using only concurrent (CC) control data or using all control data (NCC+CC controls) collected so far.}
\label{Fixed sample design}
\end{figure}

\subsection{The LOND procedure in platform trials}
In platform trials treatment arms enter and leave the trial at any time, possibly depending on previous results or available resources. Treatment arms often share one control arm (see Fig. 1) and hypothesis testing is performed for each treatment arm against the control arm. Thus for the remainder of the paper $H_i$ denotes the null hypothesis for comparing treatment arm $i$ to the data of the common control arm (indicated by the orange bar in Fig. 1). In principle the methodology presented would allow that for $H_i$ different endpoints are tested. For simplicity we assume that the same endpoint is used for each treatment-control comparison. To control the FDR of these trials with the LOND procedure, the sequence in which the hypotheses $H_1, H_2,\dots$ will be tested has to be predefined, e.g., in a platform trial by the order of entering the trial. In the following we assume equal sample sizes $n_i$ per treatment and equal allocation of observations to each treatment and the control arms, thus the sequence of hypotheses entering the study is the same as for data analysis. In case of a fixed sample design with different sample sizes $n_i$ per treatment, p-values may be available for test decision deviating from the sequence corresponding to the starting times. To control the FDR in this case it is required to maintain, e.g., the original order of the hypotheses for the calculation of the significance levels in Eq. (\ref{LOND}) \cite{Zrnic21}. Alternatively the order is fixed on the anticipated availability of the data needed for each $H_i$. It is essential that the order of the hypotheses remains independent of the observed data and may thus not be influenced by the outcome.

For a hypothesis test the control group data can be divided into two parts:
\begin{itemize}
    \item Concurrent (CC) controls: Control samples which are recruited parallel to the treatment group.
    \item Non-concurrent (NCC) controls: Control samples which are recruited from the beginning of the platform trial until the beginning of the treatment of interest.
\end{itemize}
For platform trials running over a long time period, incorporating a pooled control group of NCC and CC controls can substantially improve the power of the experiment, however, NCC controls may affect the FDR negatively in case of time trends\cite{Collignon20,Bofill21}.

\subsection{Group-sequential designs}
It is now assumed that for each treatment an interim analysis is performed after having observed a part of the pre-planned total sample size with the option of early rejection or early stopping for futility. In these group-sequential trials two sources of multiplicity must be considered:
\begin{enumerate}
    \item Adjust for the total number of hypotheses and control the online FDR with the LOND procedure (as for the design with fixed sample sizes).
    \item Adjust for the option of early rejection or stopping for futility in the interim analysis. This can be done, e.g., with spending functions\cite{Lan94}, where the significance level $\alpha_i$ is split between the first and the second stage.
\end{enumerate}
For hypothesis $H_i$ the interim analysis is performed after first stage sample size  $n_i^{(1)}$ for the treatment group and the final analysis after $n_i=n_i^{(1)}+n_i^{(2)}$ with $n_i^{(2)}$ denoting the sample size of the treatment group in the second stage. The respective first stage sample size of the control group depends on the chosen strategy of incorporating only CC or also NCC controls and the allocation ratio of controls and treatments. The p-value for the interim analysis of $H_i$ is denoted by $p_i^{(1)}$. For the final analysis the data of stage one and stage two are pooled and the p-value is given by $p_i$. The group-sequential boundaries allowing for interim analyses with the option of early rejection for individual hypotheses are determined using Lan-DeMets spending functions \cite{Lan94}. The adjusted group-sequential critical boundaries for hypothesis $H_i$ are defined by $\alpha_i^{(1)}(\alpha_i)$ for the interim analysis and $\alpha_i^{(2)}(\alpha_i)$ for the final analysis (e.g., O'Brien-Fleming \cite{OBrien79} or Pocock boundaries \cite{Pocock77}), where $\alpha_i$ is the nominal significance level dedicated to test $H_i$. This means, that $H_i$ is rejected in the interim analysis if $p_i^{(1)} \leq \alpha_i^{(1)}(\alpha_i)$ and the sample size of the treatment group in the second stage, $n_i^{(2)}$, is saved. If not, a second stage is performed subsequently and $H_i$ is rejected in the final analysis if $p_i \leq \alpha_i^{(2)}(\alpha_i)$. If additionally a stopping for futility boundary $\alpha_i^F$ is introduced, then in case of $p_i^{(1)} \geq \alpha_i^F$, $H_i$ is stopped in the interim analysis without rejecting the $H_i$ and again the sample size $n_i^{(2)}$ is saved. If $\alpha_i^{(1)}(\alpha_i) < p_i^{(1)} < \alpha_i^F$, a second stage is performed and $H_i$ is rejected in the final analysis if $p_i \leq \alpha_i^{(2)}(\alpha_i)$.

To control the online FDR for the group-sequential trial, the LOND procedure is applied for both the interim and the final analyses. $\alpha_i$ depends on the number of previously rejected hypotheses, thus, it may differ between the interim and final analysis of $H_i$. In the following 3 different LOND procedures for group-sequential designs are proposed:
\begin{itemize}
    \item Group-sequential LOND: gsLOND
    \item Group-sequential LOND exhausting updated local level alpha: gsLOND.II
    \item Group-sequential LOND updating all ongoing tests in case of any "in-between" rejections": gsLOND.III
\end{itemize}

The methods presented are straightforward for binary or continuous endpoints. For time-to-event endpoints further specifications are needed for the group-sequential procedures, which are not covered here.

\subsubsection{Group-sequential LOND (gsLOND)}
We first illustrate the challenges of a group-sequential design with the LOND procedure based on Figure \ref{Fixed sample design}b: According to LOND, the value of the nominal level for the interim analysis of treatment 3 depends on the test decision for treatment 1 and the interim decision for treatment 2. If treatment 3 proceeds to the final analysis, the level additionally may be influenced by the final analysis of treatment 2: If either treatment 2 was rejected in the interim analysis, or treatment 2 was neither rejected in the interim nor the final analysis, the sum of rejections does not change and the level for treatment 3 remains the same. If, however, treatment 2 was not rejected in the interim analysis, but indeed in the final analysis, the sum of rejections increases by one and the local level for treatment 3 has therefore to be updated and increased.

More formally, the significance level for testing $H_i$ at interim analysis is defined by

\begin{equation} 
\alpha_{i}^{I^{(1)}}=\beta_i(\sum_{j\in I_i^{(1)}} R_j^{I_i^{(1)}}+1), \label{alphainterim}
\end{equation}

and the corresponding group-sequential critical boundary is given by

$\alpha_i^{(1)}(\alpha_i^{I^{(1)}})$, where $I_i^{(1)}$ is an index set including all hypotheses $j$ (with $j<i$) where for $H_j$  already group-sequential tests have been performed before the interim analysis of $H_i$. 
$R_j^{I_i^{(1)}}=1$ for $j \in I_i^{(1)}$ if $H_j$ has been already rejected (at the interim analysis 
or at the final analysis) and it is zero, if no rejection has been performed yet. 

The level of the significance level for hypothesis $i$ may increase from interim analysis to final analysis as the sum of rejections may increase. 
For the final analysis of $H_i$, the corresponding index set is given by $I_i^{(2)}$ including all j (with $j<i$) where for $H_j$ already group-sequential tests have been performed before the final analysis of $H_i$ and the significance level is given by

\begin{equation}
\alpha_{i}^{I^{(2)}}=\beta_i(\sum_{j\in I_i^{(2)}} R_j^{I_i^{(2)}}+1).
\label{alphafinal}
\end{equation} 

The group-sequential critical boundary for the final analysis is then $\alpha_i^{(2)}(\alpha_i^{I^{(2)}}).$

If no hypotheses are rejected between interim analysis and final analysis of $H_i$,  for all $j<i$, $R_j^{I_i^{(1)}}=R_j^{I_i^{(2)}}$, and $\alpha_i^{I^{(1)}}=\alpha_i^{I^{(2)}}$.

Remark: If there was a change from $\alpha_i^{I^{(1)}}$ to  $\alpha_i^{I^{(2)}}$ (i.e., $\alpha_i^{I^{(1)}} \neq \alpha_i^{I^{(2)}}$) in the nominal level for the group-sequential test for $H_i$ due to further rejections of some $H_j$ (for $j<i$) between the interim analysis and final analysis for $H_i$, then we could formally reject $H_i$ at the final analysis not only if $p_i \leq \alpha_i^{(2)}(\alpha_i^{I^{(2)}})$, but also if only $p_i^{(1)} \leq \alpha_i^{(1)}(\alpha_i^{I^{(2)}})$ holds. The latter means that we would re-perform the test at interim analysis using an increased interim significance level. As $\alpha_i^{(1)}(\alpha_i^{I^{(1)}}) < \alpha_i^{(1)}(\alpha_i^{I^{(2)}})$ when using the same type of spending function with an increased nominal level $\alpha_i$ for the group-sequential test, we can only get additional rejections for $H_i$. However, it might be unusual to reject the null hypothesis using only part of the data in a situation where all data are available (and especially when all data suggest to retain the null hypothesis  with $p_i > \alpha_i^{(2)}(\alpha_i^{I^{(2)}})$). So we will propose a different testing strategy utilizing the alpha not used in the group-sequential testing in the situation where the nominal level alpha is increased between interim and final analysis due to other additional in-between rejections. 

\subsubsection{Group-sequential LOND exhausting updated local level alpha (gsLOND.II)}
Instead of using the same type of spending function with an increased nominal level $\alpha_i^{I^{(2)}}$, we propose to use the increment $\alpha_i^{I^{(2)}}- \alpha_i^{(1)}(\alpha_i^{I^{(1)}})$ when calculating the final level for $H_i$ in case further rejections of some $H_j$ (for $j<i$) occurred after the interim analysis of $H_i$. $\alpha_i^{(1)}(\alpha_i^{I^{(1)}})$ corresponds to the alpha already spent at the interim analysis (where the corresponding hypothesis $i$ could not be rejected) and overall the level $\alpha_i^{I^{(2)}}$ is exhausted for $H_i$. Thus if the nominal level $\alpha_i^{I^{(2)}}$ is increased, the final group-sequential plan does not correspond to the initially selected type of spending function, but exhausts fully the local nominal level as determined by the LOND procedure.

Thus for gsLOND.II, 
we extend the gsLOND procedure by modifying the group-sequential boundaries when updating the local nominal level alpha.  $\alpha_i^{I^{(1)}}$ and $\alpha_i^{I^{(2)}}$ are derived as described for the gsLOND in Eq. (\ref{alphainterim}) and (\ref{alphafinal}). However, the level for the final analysis is now modifed by recalculating the group-sequential boundary based on the increment $\alpha_i^{I^{(2)}}-\alpha_i^{(1)}(\alpha_i^{I^{(1)}})$.

\subsubsection{Group-sequential LOND - update of all ongoing tests in case of any "in-between" rejections  (gsLOND.III)}

In the group-sequential design, rejections for the hypotheses may occur non-consecutively: As illustrated in Fig. \ref{Fixed sample design}b, even if treatment 2 is still running, treatment 3 may already have been rejected in the interim analysis. However, for gsLOND, the test decisions for $H_i, H_{i+1}, \dots$ have no influence on the significance level of $H_{i-1}$.

In this section we propose a group-sequential LOND modifying the set of hypotheses which will be updated in case of any "in-between" rejections.  For gsLOND.III we propose to update the local nominal significance level for $H_i$ (which is still under investigation) not only on test decisions on $H_1,\dots,H_{i-1}$, but also on test decisions for  $H_{i+1},H_{i+2},\dots,$ if they were conducted before the (interim or final) analysis of $H_i$. Note that this does not involve re-decision of a hypothesis test if a new test decision is performed.

$$\alpha_{i}^{J^{(1)}}=\beta_i(\sum_{j\in J_i^{(1)}} R_j^{J_i^{(1)}}+1)$$ and $$\alpha_{i}^{J^{(2)}}=\beta_i(\sum_{j\in J_i^{(2)}} R_j^{J_i^{(2)}}+1)$$ denote the nominal significance levels  to calculate the group-sequential boundaries for the interim and the final test for $H_i$, respectively. $J_i^{(1)}$ and $J_i^{(2)}$ again are index sets including all hypotheses with a final test decision at the time of interim or final analysis for $H_i$. The index sets $J_i^{(1)}$ and $J_i^{(2)}$  may include all $j$ of the set $\{1,2,...,i-1,i+1,i+2, ...)$ where for $H_j$  already group-sequential tests have been performed (allowing now also for $j>i$ as well). If there were no additional rejections of other hypotheses between the interim and final analysis for $i$, then $\alpha_{i}^{J^{(1)}}=\alpha_{i}^{J^{(2)}}$, but the group-sequential boundaries may differ based on the spending function used (for more details see toy example). If the nominal significance level is increased, then the group-sequential boundaries are calculated using the initially fixed spending function as described for gsLOND.

Note that the gsLOND.III gives no formal online FDR control as it does not fulfil the condition that the local significance level $\alpha_i$ for $H_i$ is a function of the decisions for $H_1,\dots,H_{i-1}$ only, whereby in the original LOND procedure a pre-fixed order has to be used. But by not exhausting the nominal level alpha when updating the group-sequential boundaries, this might compensate for in principle violating the predefined order in LOND.

\subsubsection{gsLOND.II.III}
The two modifications gsLOND.II and gsLOND.III may also be combined, e.g. if $H_{i-1}$ and $H_{i+1}$ both were rejected between the interim and the final analysis of $H_i$. Thus, for the final analysis of $H_i$ first the group-sequential boundaries are updated to exhaust the local nominal level (gsLOND.II) due to rejection of $H_{i-1}$ only after the interim analysis of $H_i$ and second the updated sum of rejections is applied for the calculation of the level (gsLOND.III). We will illustrate this in the toy example.

\begin{table}[h]
\centering
\caption{Toy example. Nominal (group-sequential) significance levels for the hypotheses in case of 0, 1 or 2 previous rejections ($\sum R$) for the LOND and the gsLOND design for $\alpha=0.05$ (one-sided), a total number of $K=3$ hypotheses, and equal $\beta_i=0.0167$. \label{T2}}
\begin{tabular}{lccccc}
\hline
&& \multicolumn{2}{c}{PO} & \multicolumn{2}{c}{OBF}\\ 
$\sum R$ & LOND: $\alpha_i$ & \multicolumn{2}{c}{gsLOND} &  \multicolumn{2}{c}{gsLOND}\\ 
&& stage 1 & stage 2 & stage 1 & stage 2\\ \hline
0 &0.0167  & 0.0103 & 0.0089  & 0.0007 & 0.0164 \\
1 & 0.0334 & 0.0207 & 0.0190 & 0.0026 & 0.0325 \\
2 & 0.05   & 0.0310 & 0.0297 & 0.0056 & 0.0482 \\
\hline
\end{tabular}\\[10pt]
\end{table}

\subsection{Toy example}
To illustrate the group-sequential LOND procedures we present a toy example and derive the appropriate critical boundaries: The number of hypotheses is set to $K=3$ and the significance level $\alpha=0.05$ (one-sided test) is distributed equally among the 3 hypotheses, i.e., $\beta_i=0.0167$, $i=1,2,3$ (note that $\alpha=0.05$ in the toy example, whereas in the simulation studies below, $\alpha=0.025$). This means, for this toy example, we fix the total number of hypotheses straight at the beginning of the trial and do not give the option of including additional hypotheses. An ordering of the interim and final analyses is considered as in Fig. \ref{Fixed sample design} b). For this scenario, Table \ref{T2} shows the critical boundaries for the fixed sample and the group-sequential design for the three hypotheses for the LOND and the gsLOND procedures for the Pocock\cite{Pocock77} (PO) and O'Brien- Fleming\cite{OBrien79} (OBF) type $\alpha$-spending design in case of 0, 1, or 2 previous rejections. Only for this special case with equal $\beta_i$, the boundaries for $H_i$ depend on $i$ through the number of previous rejections of hypotheses $H_j$ (with $j<i$).   

Table \ref{T3} shows the group-sequential critical boundaries for gsLOND, gsLOND.II, gsLOND.III, and gsLOND.II.III for $H_2$ for all possible outcomes of the tests for $H_1$ and $H_3$ in the interim or in the final analyses (PO design). Similarly, Table \ref{T4} shows the boundaries for $H_3$. If the relevant sum of rejections does not change between the interim and the final analysis, the gsLOND procedure for the final analysis has the same boundaries as the gsLOND.II and gsLOND.III procedures. With relevant sum it is meant, that for updating the level of $H_2$, only rejections of $H_1$ will matter for the procedures gsLOND and gsLOND.II, whereas for gsLOND.III and gsLOND.II.III also rejections of $H_3$ will matter. The group-sequential critical boundaries for gsLOND and gsLOND.III are calculated with the R-package\cite{R} ldbounds\cite{ldbounds}, for the other procedures rpact\cite{rpact} is applied.

\paragraph{gsLOND.II} In Table \ref{T4} the gsLOND.II procedure can be applied for the calculation of the group-sequential boundaries for $H_3$ if $H_2$ has been rejected in the final but not in the interim analysis: E.g, in line 3, it is assumed that $H_1$ has been retained and $H_2$ has been rejected in the interim analysis and therefore, $\alpha_3^{I^{(1)}}=\alpha_3^{I^{(2)}}=\beta_2 (1+1)=0.0334$, $\alpha_3^{(1)}=0.02076$, and $\alpha_3^{(2)}=0.0190$. If, however, $H_2$ is retained in the interim analysis and then rejected in the final analysis, this additional rejection cannot be applied for the calculation of the level for the interim analysis of $H_3$ of gsLOND due to a time overlap (final analysis of $H_2$ after interim analysis of $H_3$) and thus $\alpha_3^{I^{(1)}}$ was only set to 0.0167 instead of 0.0334. With the gsLOND.II method this "loss" is translated to the final analysis of $H_3$ by exhausting the local level and increasing $\alpha_3^{(2)}$ to 0.0279 instead of 0.0190.

\paragraph{gsLOND.III} In line 4 in Table \ref{T3} it is assumed that $H_1$ has been rejected in the interim analysis and thus for the interim analysis of $H_2$, $\alpha_2^{I^{(1)}}=\beta_2 (1+1)=0.334$. The corresponding group-sequential boundary is $\alpha_2^{(1)}=0.0207$. If $H_3$ is not rejected, $\alpha_2^{I^{(2)}}=\alpha_2^{I^{(1)}}$ and $\alpha_2^{(1)}=0.0190$. If, however, $H_3$ was rejected in the interim analysis (Table \ref{T3}, line 5) and this decision was made before the final analysis of $H_2$, for the gsLOND.III,  $\alpha_2^{I^{(2)}}=\beta_2 (2+1)=0.05$ and thus the group-sequential level can be increased to $\alpha_2^{(2)}=0.0297$.

\begin{table}[h]
\centering
\caption{Toy example. Group-sequential boundaries for $H_2$ with a sequence of hypotheses according to Fig.\ref{Fixed sample design}b with $\alpha=0.05$ (one-sided), $K=3$, and PO design. Scenarios where gsLOND.II, gsLOND.III and/or gsLOND.II.III differ from gsLOND are marked in bold. \label{T3}}
\begin{tabular}{cccccccccc}
\hline
&&\multicolumn{2}{c}{gsLOND} & gsLOND.II & gsLOND.III & gsLOND.II.III
\\
$H_1$& 
$H_3$ &stage 1& stage 2  &  stage 2 & stage 2 & stage 2
\\\hline
retain & 
retain & 0.0103 & 0.0089 & 0.0089& 0.0089& 0.0089\\
\textbf{retain} & 
\textbf{reject interim} & 0.0103 & 0.0089 & 0.0089  &\textbf{0.0190}& 0.0190\\
retain & 
reject final & 0.0103 & 0.0089 & 0.0089& 0.0089& 0.0089\\
reject interim & 
retain & 0.0207 & 0.0190 & 0.0190& 0.0190& 0.0190\\
\textbf{reject interim} &
\textbf{reject interim}  & 0.0207 & 0.0190 & 0.0190& \textbf{0.0297} &0.0297\\
reject interim & 
reject final & 0.0207 & 0.0190 & 0.0190& 0.0190& 0.0190\\
\textbf{reject final} &
\textbf{retain }  & 0.0103 & 0.0190 & \textbf{0.0279}& 0.0190& 0.0279\\
\textbf{reject final} &
\textbf{reject interim}  & 0.0103 & 0.0190 &\textbf{0.0279}& \textbf{0.0297} & \textbf{0.0459}\\
\textbf{reject final} &
\textbf{reject final}  & 0.0103 & 0.0190 & \textbf{0.0279} & 0.0190& 0.0279\\
\hline
\end{tabular}\\[10pt]
\end{table}

\begin{table}[h]
\centering
\caption{Toy example. Group-sequential boundaries for $H_3$ with a sequence of hypotheses according to order of Fig. \ref{Fixed sample design}b with $\alpha=0.05$ (one-sided), $K=3$, and PO design. Scenarios where gsLOND.II differs from gsLOND are marked in bold. \label{T4}}
\begin{tabular}{cccccccccc}
\hline
$H_1$ & $H_2$ & 
\multicolumn{2}{c}{gsLOND}  & gsLOND.II 
\\
&&stage 1& stage 2 & stage 2 
\\\hline
retain & retain &
0.0103 & 0.0089 &  0.0089
\\
reject &retain & 
0.0207 & 0.0190 &  0.0190
\\
retain & reject interim & 
0.0207 & 0.0190 &  0.0190
\\
\textbf{retain} & \textbf{reject final}  & 
\textbf{0.0103} & \textbf{0.0190} & \textbf{0.0279}\\
reject & reject interim & 
0.0310 & 0.0297 & 0.0297\\
\textbf{reject} & \textbf{reject final} & 
\textbf{0.0207} & \textbf{0.0297} & \textbf{0.0389}\\
\hline
\end{tabular}\\[10pt]
\end{table}

\section{Results}
\subsection{Simulation methods}

Simulation studies were performed to compare the LOND procedures for different scenarios for fixed and group-sequential designs with regard to average power, defined as the proportion of rejected alternatives among all alternatives, actual FDR and the average percentage amount of saved sample size of the group-sequential design compared to the fixed sample design.
The following procedures are considered:
\begin{itemize}
    \item Fixed sample design: LOND procedure
    \item Group-sequential design: gsLOND, gsLOND.II, gsLOND.III, level-$\alpha$ test, Bonferroni 
\end{itemize}

For the level-$\alpha$ test, no adjustment is performed for multiple hypothesis testing, only for the interim analyses. Thus for each $H_i$, $\alpha_i=\alpha$ and group-sequential critical boundaries are derived according to the OBF or the PO design, respectively. In contrast, for the group-sequential Bonferroni procedure the significance level $\alpha_i$ for each hypothesis is adjusted according to the Bonferroni method and used as nominal level alpha for the group-sequential plan. Even though the total number of hypotheses $K$ is unknown, in the simulations we set $\alpha_i=\alpha/K$ (best case scenario). 

In each scenario we compare either $K=10$ or $K=100$ treatment groups with one single control group. As shown in Fig. \ref{Fixed sample design}, recruitment of the control group starts at the beginning of the platform trial and patients are recruited during the whole observation period. Treatments start at different times and run parallel to the control group, depending on the individual start. We interpret the index of the control patients as the measure for time and assume that control patients enter the trial consecutively. We further assume that the distribution of patients to control and treatments running in parallel is equal for all arms, i.e. if two treatment arms 1 and 2 are running, the distribution is 1:1:1 for control, treatment 1, and treatment 2.   
The sample size for each treatment is $n=50$ and in case of an interim analysis it is equally divided between two stages of the treatment group, $n^{(1)}=n^{(2)}=25$ (which can easily be extended to varying per-hypothesis sample sizes $n_i$, $n_i^{(1)}$, $n_i^{(2)}$). For the first part of simulations we assume that after every $n_\Delta=20$ control patients, a new treatment arm starts. 
In case of CC controls only, for each hypothesis test, the observations for the control and the corresponding treatment run parallel and both groups have equal sample sizes. 
In the case of NCC controls, the control group additionally contains all control observations collected so far 
and thus has a potentially larger first stage sample size. The observations for a treatment group are the same in the CC and in the NCC scenario.

Normally distributed observations with $N(\Delta,1)$ were simulated with $\Delta=0$ for the control group and treatment data from the set of true null hypotheses and $\Delta=0.6$ for treatments of true alternative hypotheses with a proportion of true null hypotheses of $\pi_0$. For each treatment we considered the one-sided null hypothesis $$H_{0i}: \mu_i \leq 0\ \textrm{ versus }\ H_{1i}: \mu_i>0$$ for the mean of the observations. Two-sample t-tests were performed and corresponding one-sided p-values were derived. 
The significance level for the online FDR procedures was set to $\alpha=0.025$ and stopping for futility was performed in the interim analysis if $p_i^{(1)}>\alpha^F=0.5$.
The values of $\beta_j$ for the LOND procedure were first calculated as proposed by, e.g,  Javanmard and Montanari (2018)\cite{Javanmard18}: \begin{equation} \label{eqn:beta} \beta_j=C\alpha \frac{\log(\max(j,2))}{j\exp(\sqrt{\log j})} \end{equation} with $C=0.07720838$ resulting in a decreasing sequence of numbers with, e.g., $\beta_1=0.00134$, $\beta_2=0.00029$, $\beta_3=0.00025$. Setting an upper bound $N$ on the total number of hypotheses as suggested by Robertson and Wason (2018)\cite{Robertson18} such that $\sum_{i=1}^N \beta_i=\alpha$, leads to larger values of $\beta_i$ and consequently larger significance levels. E.g., for $N=1000$, $\beta_1=0.00446$, $\beta_2=0.00099$, $\beta_3=0.00083$.

\paragraph{Order of alternatives:} In the simulations we considered three different orders of alternative hypotheses (see supplemental material, Fig.1). 
\begin{itemize}
\item Random order: In the simulations for each hypothesis the probability for a null hypothesis is $\pi_0$. Thus for individual simulation steps, the number of alternatives may differ.
\item Alternatives first: The simulated platform trial starts with $m_1=K-\pi_0 K$ alternatives followed by $K-m_1$ true null hypotheses. 
\item Alternatives last: The simulated platform trial starts with $K-m_1$ true null hypotheses followed by $m_1$ alternative hypotheses. 
\end{itemize}

If not specified otherwise, group-sequential critical boundaries were derived according to the O'Brien-Fleming (OBF) type alpha-spending design (R-packages\cite{R} ldbounds\cite{ldbounds} or rpact\cite{rpact}). Note that in the simulations we used the same weights for the spending function for NCC controls as for CC controls and thus the same group-sequential boundaries. Thus the procedure for NCC controls becomes conservative as it does not fully use the actual correlation.  The calculation of the efficacy critical boundaries does not incorporate the futility stopping, thus the futility threshold is non-binding \cite{bretz2009adaptive}. All simulations were conducted using the R program \cite{R}, for each scenario at least 5000 simulation runs were performed.
Note that in all considered simulation scenarios, the FDR is maintained at one-sided level $\alpha=0.025$ (exception: level-$\alpha$ test). Details can be found in the supplemental material.

\subsection{Comparison of LOND methods}
\subsubsection{CC controls}
\begin{figure*}
\centering
\includegraphics[width=12cm]{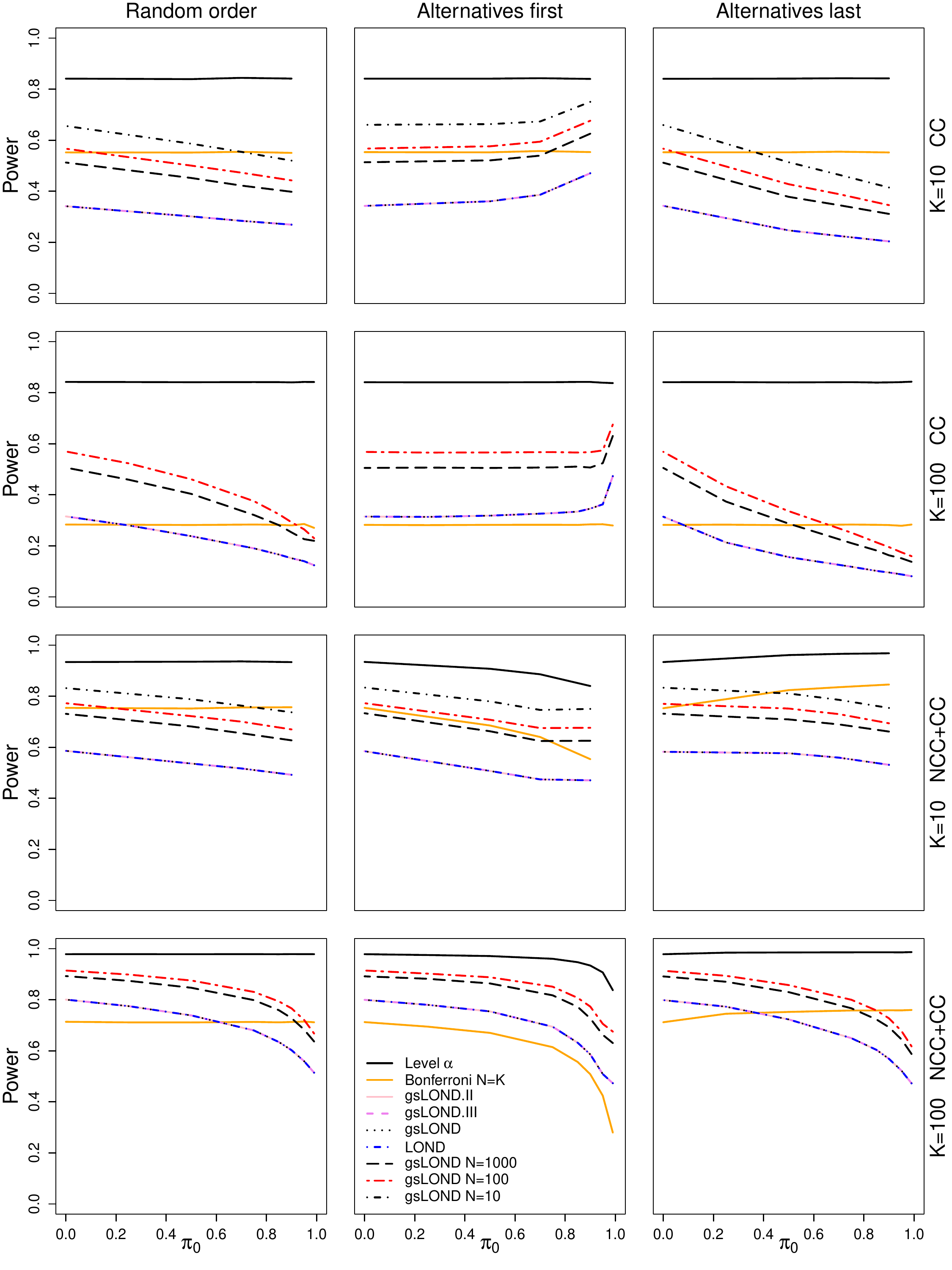}
\caption{Power for scenarios with CC (rows 1 and 2) and NCC+CC controls (rows 3 and 4) as a function of $\pi_0$ for the level-$\alpha$ and the Bonferroni procedure (with $N=K$) and the four LOND procedures. OBF design, $N=\{10,100,1000,\infty\}$, $\Delta=0.6$, $\alpha=0.025$, $\alpha^F=0.5$. The four LOND procedures can hardly be distinguished as the power values are very similar. Thus, for $N<\infty$, only the power for gsLOND is depicted.}
\label{fig:01}
\end{figure*}

Overall power values as a function of $\pi_0$ for a simulation trial with CC control data for the level-$\alpha$ test, the Bonferroni procedure and the 4 LOND procedures for unlimited number of hypotheses $N$ are shown in Fig. \ref{fig:01} (rows 1-2). Additionally, gsLOND for $N \in \{10,100,1000\}$ (for $K=100$, only $N\in \{100,1000\}$) was considered.
No differences in the power curves can be observed between the 4 LOND procedures with unlimited $N$, only in the third decimal place the power values differ. The reason is that for the gsLOND procedures the OBF group-sequential boundaries for interim analyses are rather low, early rejections thus only occur for large effects. However, for these cases, also the fixed sample LOND procedure leads to a rejection and the advantage of the gsLOND is only comprehensible in the amount of saved sample size. Below we will assume a simulation scenario, where additional treatments are included in case of early stopping in the interim analysis and we will show that the number of rejected alternatives increases for gsLOND procedures compared to the fixed sample design.

Setting $N=1000$, the power of gsLOND increases by more than 10 percentage points. It further increases for $N=100$ or $N=10$ (for $K=10$ only), but the improvement is less pronounced. Note that again nearly equal power values are observed for LOND, gsLOND, gsLOND.II, and gsLOND.III with $N \in \{10,100,1000\}$ (data not shown).

The power of the LOND procedures depends on the order of alternatives: For "alternatives first", the power increases with the proportion of null hypotheses $\pi_0$, the first alternative has the largest level and thus the largest power, additional alternatives only decrease the power on average. For "alternatives last" the situation is reversed, LOND procedures have decreasing power values for increasing $\pi_0$, because for "late" alternatives with large $i$ the values of $\beta_i$ are only low and hardly any rejections have been performed before. Here the Bonferroni procedure (best case scenario) has even higher power than the comparable gsLOND procedure with $N=K$. The level-$\alpha$ and the Bonferroni procedure are not influenced by the order of alternatives as the significance levels are constant.

\subsubsection{NCC+CC controls}
The same simulations as described above were repeated for NCC+CC controls (Fig. \ref{fig:01}, rows 3 and 4). 
As for the scenarios with CC controls, no differences in the power curves can be observed between the 4 LOND procedures and the power values very much depend on the value of $N$. For the scenario $K=10$ and "alternatives first", the power is not monotonous in $\pi_0$. The reason is the trade off between decreasing level and increasing sample size for the NCC control data. 

The power for the level-$\alpha$ and the Bonferroni procedure now also depends on the distribution of alternatives due to the composition of the control group and is therefore not constant: E.g., comparing scenarios "alternatives first" and "alternatives last" for large $\pi_0$, the number of observations in the NCC+CC control group for the "early" true alternative(s) is much lower and thus also power is decreased.

\begin{figure*}
\centering
\includegraphics[width=12cm]{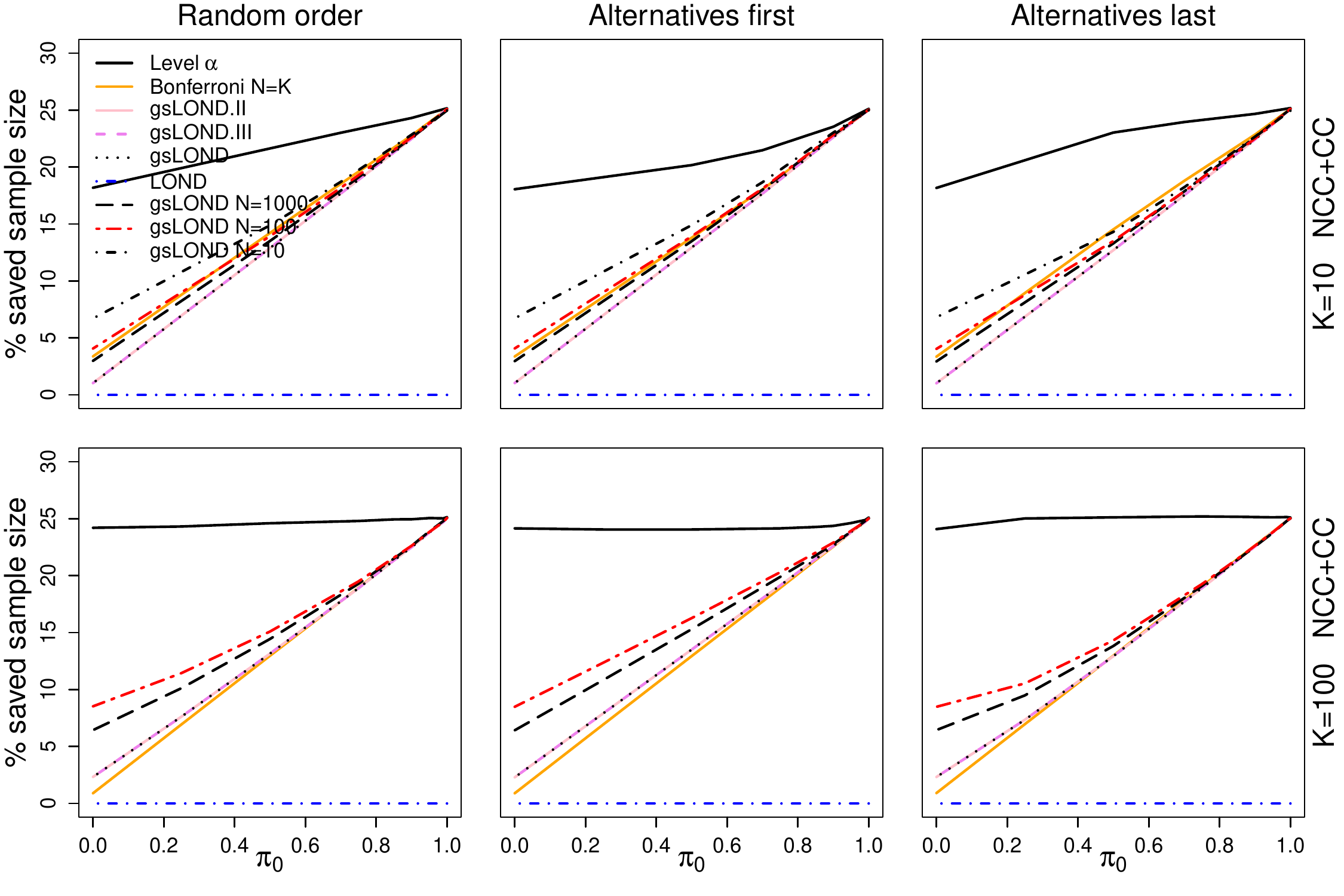}
\caption{\% saved sample size for NCC+CC controls as a function of $\pi_0$ for the level-$\alpha$ and the Bonferroni procedure (with $N=K$), and the four LOND procedures. OBF design, $N=\{10,100,1000,\infty\}$, $\Delta=0.6$, $\alpha=0.025$, $\alpha^F=0.5$. The four LOND procedures can hardly be distinguished as the power values are very similar. Thus, for $N<\infty$, only the power for gsLOND is depicted.}
\label{fig:01_saved}
\end{figure*}

The advantage of the group-sequential LOND procedures in comparison to the fixed sample LOND can be seen in the average percentage amount of saved sample size in Fig. \ref{fig:01_saved} (NCC+CC controls). The average percentage saved sample size is defined as the proportion of actually saved treatment observations in stage 2 due to early stopping for efficacy or futility in the interim analysis. For $K=10$, the maximum total number of treatment observations in stage 2 is given by $Kn^{(2)}=250$ and for $K=100$ by $2500$. For small $\pi_0$ the percentage saved sample size is low as only hypotheses with a very small p-value are rejected in the interim analysis, but it increases for decreasing $N$. With increasing $\pi_0$, more sample size can be saved due to early stopping for futility in the interim analyses. For $\pi_0=1$, on average 25\% can be saved for $\alpha^F=0.5$. Simulated FDR values and \% saved sample size for CC controls can be found in the supplemental material (Fig. 2-4).

\subsubsection{Direct comparison of concurrent and non-concurrent controls}
\begin{figure*}
\centering
\includegraphics[width=12cm]{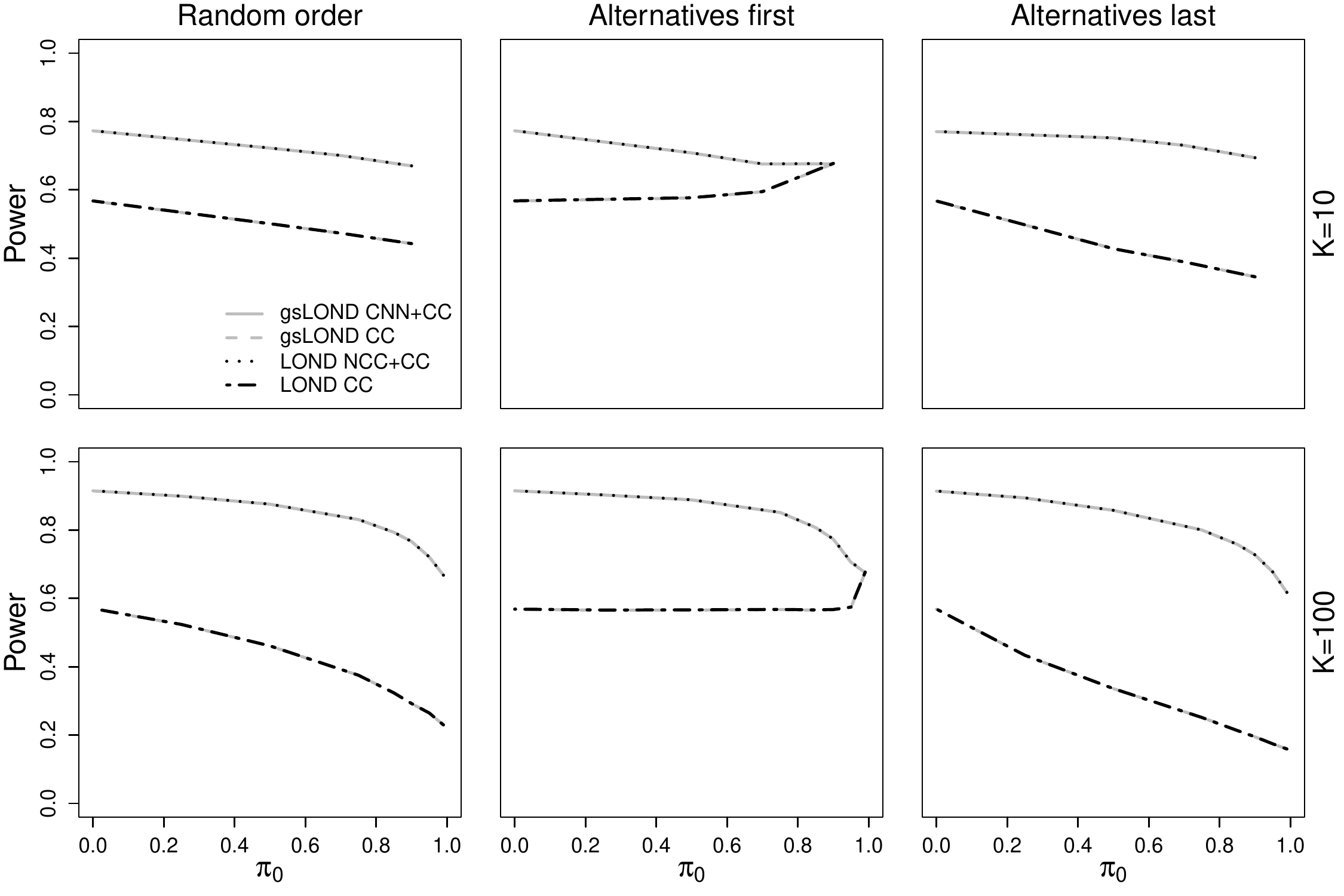}
\caption{Power for CC versus all (NCC+CC) controls as a function of $\pi_0$ for LOND and gsLOND (results for gsLOND.II and gsLOND.III are not depicted due to nearly identical power values); OBF design, $N=100$, $\Delta=0.6$, $\alpha=0.025$, $\alpha^F=0.5$.}
\label{fig:2}
\end{figure*}

Fig. \ref{fig:2} compares NCC+CC controls versus CC controls for LOND and gsLOND procedures with upper bound $N=100$. As expected, depending on the distribution of the alternatives, the inclusion of all controls recruited so far increases the power values, particularly if true alternatives arise at the end of a platform trial and the sample size of the control group is large. Note again that the power values of gsLOND and LOND are equal. The values of the actual FDR for the CC and the NCC+CC control data are comparable (see Fig.6 in the supplemental material).

\subsection{Distribution of significance level}
\begin{figure*}
\centering
\includegraphics[width=12cm]{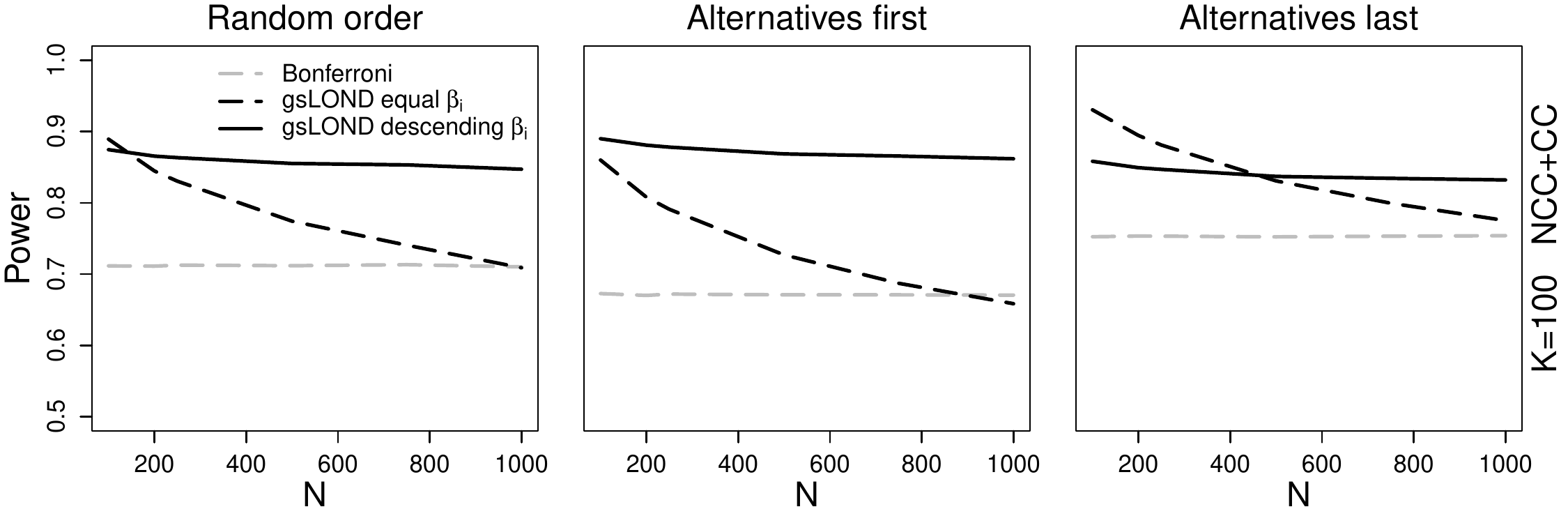}
\caption{Power of gsLOND (NCC+CC controls) for two distributions of the significance level as a function of $N$ and Bonferroni for $N=K$. OBF design, $\Delta=0.6$, $\pi_0=0.5$, $K=100$, $\alpha=0.025$, $\alpha^F=0.5$.}
\label{fig:4a}
\end{figure*}

In the previous simulations the values of $\beta_i$ were calculated according to Eq.\ref{eqn:beta} and possibly adjusted by the upper bound $N$ and thus descending with increasing number of hypotheses ("descending $\beta_i$"). We additionally consider that significance level $\alpha$ is equally distributed among all (potential) hypotheses ("equal $\beta_i$"), i.e., $N$ has to be prefixed and $\beta_i=\alpha/N$. In Fig. \ref{fig:4a} the impact of "descending $\beta_i$" and "equal $\beta_i$" on the power of the group-sequential LOND is investigated as a function of $N$ (for $K=100$, $\pi_0=0.5$, NCC+CC controls, OBF design; results for CC controls can be found in Fig. 7 in the supplemental material). The results of the procedures depend strongly on the value of $N$:  In case of random order, only for $N$ very close to $K$ "equal $\beta_i$" is superior, in case of alternatives last, "equal $\beta_i$" is superior for $N<500$. If, e.g., $N=1000$ and $K=100$, the significance level for "equal $\beta_i$" is $\beta_i=0.000025$  for each hypothesis. Until $H_{138}$, the nominal level of the "descending $\beta_i$" method and thus the power is higher, only as of $\beta_{138}$ the descending level is lower. However, for $N=100$, as of $\beta_{14}$ the "equal $\beta_i$" level is larger. The Bonferroni procedure, adjusted by the true number of hypotheses $K$ has a lower power for most scenarios, only for large $N$ it has similar power values as "equal $\beta_i$".

\subsection{OBF versus PO design}
\begin{figure*}
\centering
\includegraphics[width=8cm]{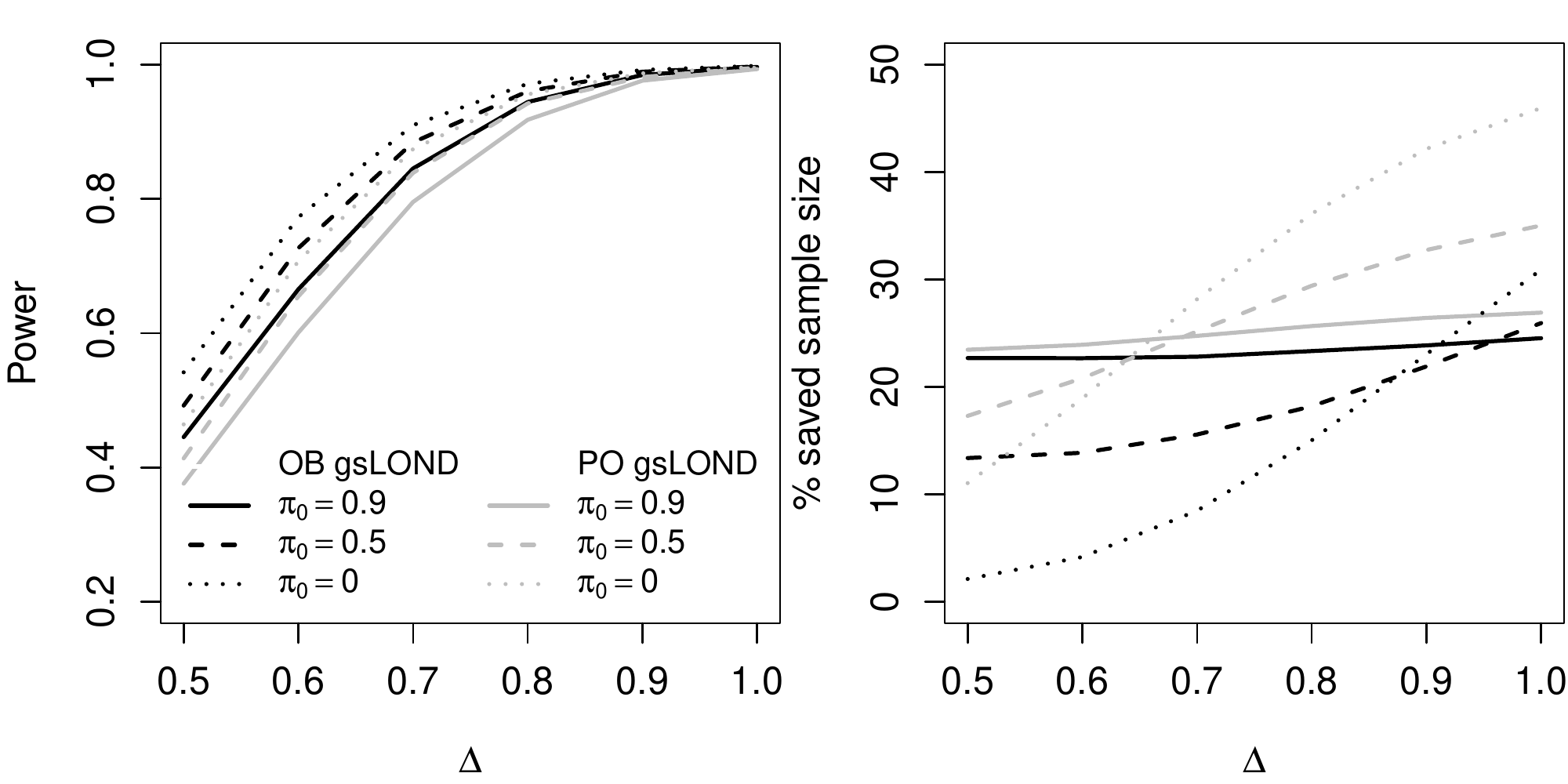}
\caption{Power and \% saved sample size for OBF and PO designs of gsLOND as a function of $\Delta$ for $\pi_0=\{0,0.5,0.9\}$ and $K=10$, $N=100$, $\alpha=0.025$, $\alpha^F=0.5$, NCC+CC controls, random order of alternatives.}
\label{fig:5}
\end{figure*}

The OBF design only distributes a small amount of the significance level to the interim analysis and retains a large amount for the final analysis, whereas the PO type $\alpha$-spending design more or less distributes the level equally between interim and final analysis. Fig \ref{fig:5} shows the power values for the two designs as a function of the effect size for $\pi_0=\{0,0.5,0.9\}$ and $K=10$, NCC+CC controls (for CC controls see Fig. 8 in the supplemental material). In terms of power, the OBF design of the gsLOND procedure is always superior compared to the PO design when using the same sample sizes in both designs. This is a well-known feature in group-sequential designs. To achieve similar power, Jennison and Turnbull\cite{Jennison00} have shown that when testing a single hypothesis, a larger maximum sample size would be needed for PO designs compared to fixed sample or OBF designs. Within the PO scenarios, the gsLOND procedure has slightly less power than gsLOND.II or seq.LOND.III (see Fig. 9 in the supplemental material). For larger $K$ the difference becomes negligible. 
However, due to spending more alpha at interim, the comparison of the saved sample size reveals that the amount of saved sample size of the PO design is always higher than for the OBF design, for low $\pi_0$, this difference in some scenarios is more than doubled. 

\subsection{Inclusion of additional treatments for stopping in the interim analysis}
Now we assume that the budget of the whole platform trial, i.e., the total number of observations, is fixed with budget $B=\sum_{i=1}^{K_0} n_i+C$ where $K_0$ denotes the number of initially planned trials and $C$ the number of pre-planned controls. If a treatment arm is dropped early due to efficacy or futility and sample size may be saved, an additional treatment arm can be included immediately if the total budget $B$ is not yet exhausted. The ordering of the hypotheses is based on the order of entrance into the platform. For the effect size $\Delta$ of the alternatives and the value of $\pi_0$ several assumptions are made. First, it is assumed that $\pi_0$ and the effect size for the alternatives remain constant through the trial, also for added treatments (Scenario 1). Second, distributed effect sizes are assumed with $\Delta=\{0.4,0.8,1.2\}$ randomly selected (scenario 2). In scenario 3, $\pi_0$ decreases by 1/80 for each new treatment and therefor alternative hypotheses become more likely, whereas in scenario 4 $\pi_0$ remains constant, but the value of the effect size $\Delta$ increases to $\Delta=1$ for the additional alternative hypothesis.

\begin{figure*}
\centering
\includegraphics[width=8cm]{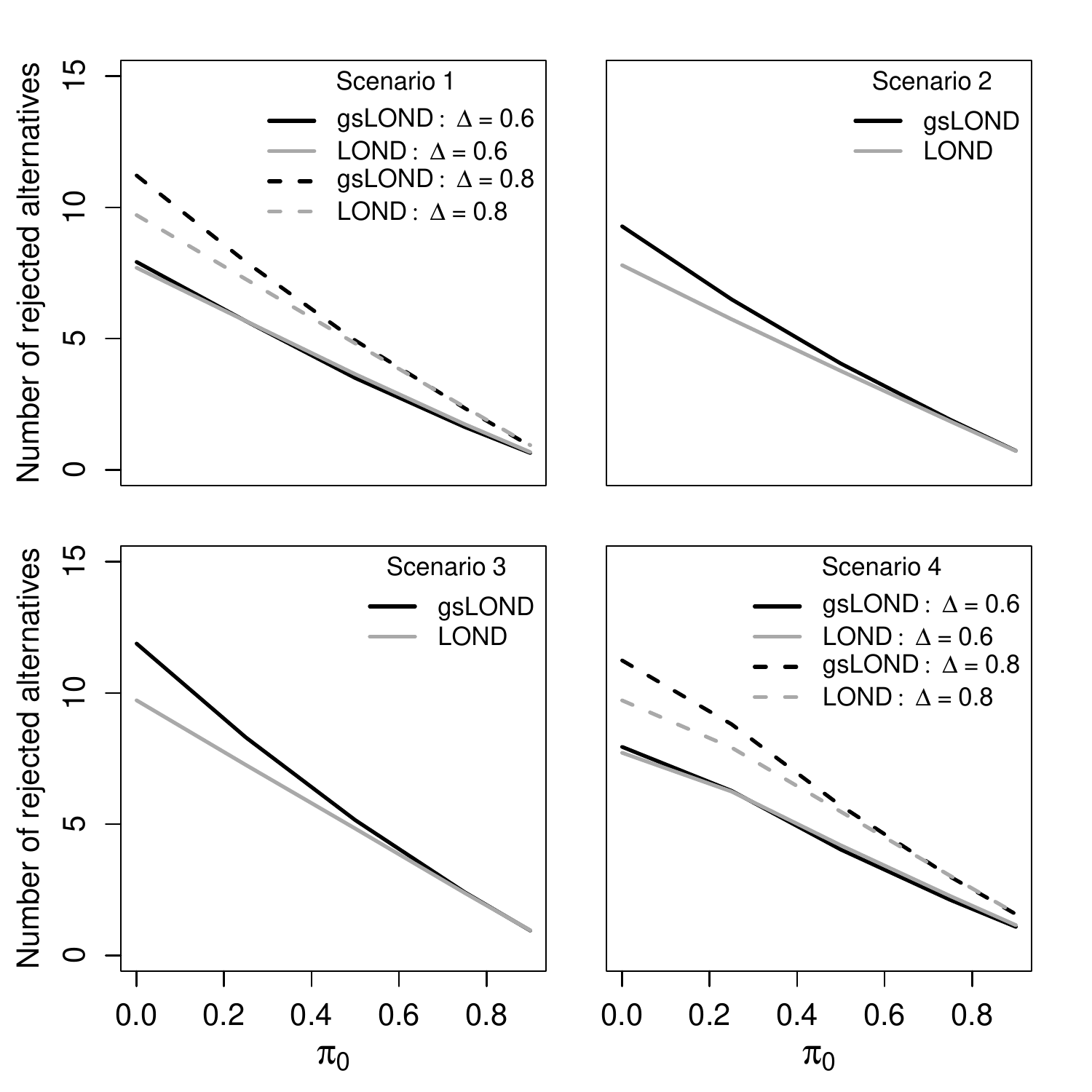}
\caption{Comparison of numbers of rejected alternatives for platform trials with a fixed budget (NCC+CC controls) as a function of $\pi_0$ for LOND and gsLOND. If a treatment arm is stopped early, an additional treatment is included. Scenario 1: $\pi_0$ and effect size remain constant, $\Delta=0.6$ and 0.8. Scenario 2: distributed effect sizes of alternatives, $\Delta=\{0.4, 0.8, 1.2\}$. Scenario 3: $\pi_0$ decreases by 1/80 for each new treatment. Scenario 4: $\Delta$ increases to $\Delta=1$ for additional alternative hypotheses. $K_0=10$, $N=100$, $\alpha=0.025$, $\alpha^F=0.5$, OBF design.}
\label{fig:6}
\end{figure*}

In the simulations we compare the LOND with the gsLOND procedure regarding the total number of rejected alternatives with $\alpha=0.05$, $\alpha^{F}=0.5$, $K=10$, $N=100$, NCC+CC controls and OBF design. Results are shown in Fig.\ref{fig:6}. In scenario 1, for small effect size $\Delta=0.6$, the procedures are rather similar, only for larger $\Delta=0.8$ and smaller $\pi_0$ the gsLOND procedures are superior compared to the LOND procedures. This effect is even more pronounced for the PO design (see supplemental material, Fig.11). For Scenarios 2-4 gsLOND is superior to LOND, only for Scenario 4 with low $\Delta=0.6$ the LOND procedure rejects a larger number of alternatives if $\pi_0$ is large.

\section{Discussion and Conclusions}
In this manuscript, the LOND procedure to control the online FDR in platform trials was examined. We showed how group-sequential methods have to be modified to allow for interim analyses both for efficacy and futility. Extensive simulation studies were performed and we observed that the proposed gsLOND, gsLOND.II, gsLOND.III and LOND with fixed sample sizes have nearly identical power values, however, a large amount of total sample size may be saved with group-sequential approaches either due to early stopping for futility or for efficacy in the interim analyses. In comparison with a group-sequential Bonferroni approach, however, depending on the value of $N$, the online FDR procedures are more powerful in many scenarios and save more sample size in most scenarios.

Setting an upper bound on the number of hypotheses has a substantial impact on the power of the LOND procedures and is thus recommended. In the simulations a large bound of $N=1000$ for an actual number of, e.g., $K=10$ hypotheses increases the power strikingly; if the bound is further decreased, the power increases even more.

For all simulation scenarios we considered positively dependent test statistics and we found no inflation of the FDR level for gsLOND, gsLOND.II, gsLOND.III, and, as expected, for LOND with fixed sample design, also under the global null hypothesis for $\pi_0=1$ (see supplemental material). A key aspect of all LOND procedures is that the hypotheses have to be ordered independently of the data to ensure a proper control of the FDR rate when updating the testing procedures due to rejections. In platform trials a natural ordering could be based on the order of entrance into the platform. However, as shown in the paper this ordering might be violated when we allow for unequal sample sizes and interim analyses. As shown by simulations the proposed procedure sufficiently controls the FDR rate, though there is no formal proof yet. Only for applying $\beta_i^{\textrm{DEP}}=\beta_i/(\sum_{j=1}^i 1/j)$ instead of $\beta_i$ in the LOND procedure, LOND also controls the FDR for general dependency of p-values\cite{Robertson18} (according to the Benjamini-Yekutieli procedure \cite{Benjamini01}) and thus also for the group-sequential design. However, we did not consider this modification in our simulations due to its conservative performance.

Simulations were performed for CC controls only and for NCC+CC controls. Also when including NCC control data, we found no inflation of the FDR level. As expected the inclusion of all so far observed controls leads to higher power values in the simulation studies, particularly for "late" hypotheses in platform trials running for a long time period ("alternatives last").  Nevertheless, the decision for or against using NCC controls must be reached for each platform trial individually, as bias may be introduced in case of time trends and the FDR control of the procedure may be negatively affected by the inclusion of NCC controls. Time trends may be caused by changes in study population over time, e.g., due to a change in standard of care \cite{Collignon20,Bofill21}.

Platform trials are a new concept for clinical study design with potentially exploratory and confirmatory aims. Currently there exists no consensus and many open issues on the type of multiplicity control\cite{Wason21,Bretz20,Collignon20}. On the one hand a rather strict adjustment is postulated by control of the FWER, defined as the probability of at least one Type I error, to prevent false positive decisions. On the other hand, no adjustment for multiplicity is recommended as platform trials are considered as a collection of independent aims. The control of the (online) FDR reflects a compromise between no adjustment and the conservative FWER adjustment \cite{Wason21,Robertson21}. The FDR is equivalent to the FWER in case all null hypotheses are true, however, it is less conservative for a positive number of false null hypotheses as false rejections are allowed as long as its expected proportion among all rejections is maintained at the pre-specified level. Thus the power of the platform trial can be increased if FDR instead of FWER control is applied.

Part of future work is also how to use the already accumulated data in an on-going platform trial to specify design aspects of new treatment arms. E.g., to reassess power and sample size based on the nominal level actually available at the start of a new arm. Furthermore, conditional power arguments may be used for sample size reassessment at interim analyses \cite{bauer2006reassessment}.

\section{Supplemental Material}
Supplementary document is available with addtional results from the simulations.

\section{Funding}
EU-PEARL (EU Patient-cEntric clinicAl tRial pLatforms) project has received funding from the Innovative Medicines Initiative (IMI) 2 Joint Undertaking (JU) under grant agreement No 853966. This Joint Undertaking receives support from the European Union’s Horizon 2020 research and innovation programme and EFPIA and Children’s Tumor Foundation, Global Alliance for TB Drug Development non-profit organisation, Springworks Therapeutics Inc. This publication reflects the authors’ views. Neither IMI nor the European Union, EFPIA, or any Associated Partners are responsible for any use that may be made of the information contained herein.

\section{Declaration of Conflicting Interests}
The Authors declare that there is no conflict of interest.

\bibliographystyle{ieeetr}
\bibliography{imsrefs_bioinformatics.bib}

\end{document}


\title{Supplemental material: Online control of the False Discovery Rate in in group-sequential platform trials}

\author{Sonja Zehetmayer\footnote{sonja.zehetmayer@meduniwien.ac.at}, Martin Posch, Franz Koenig}

\date{Center for Medical Statistics, Informatics and Intelligent Systems, Medical University of Vienna, Austria
}

\maketitle
\section{Order of alternatives}

\begin{figure}[h!]
\centering
\includegraphics[width=15cm]{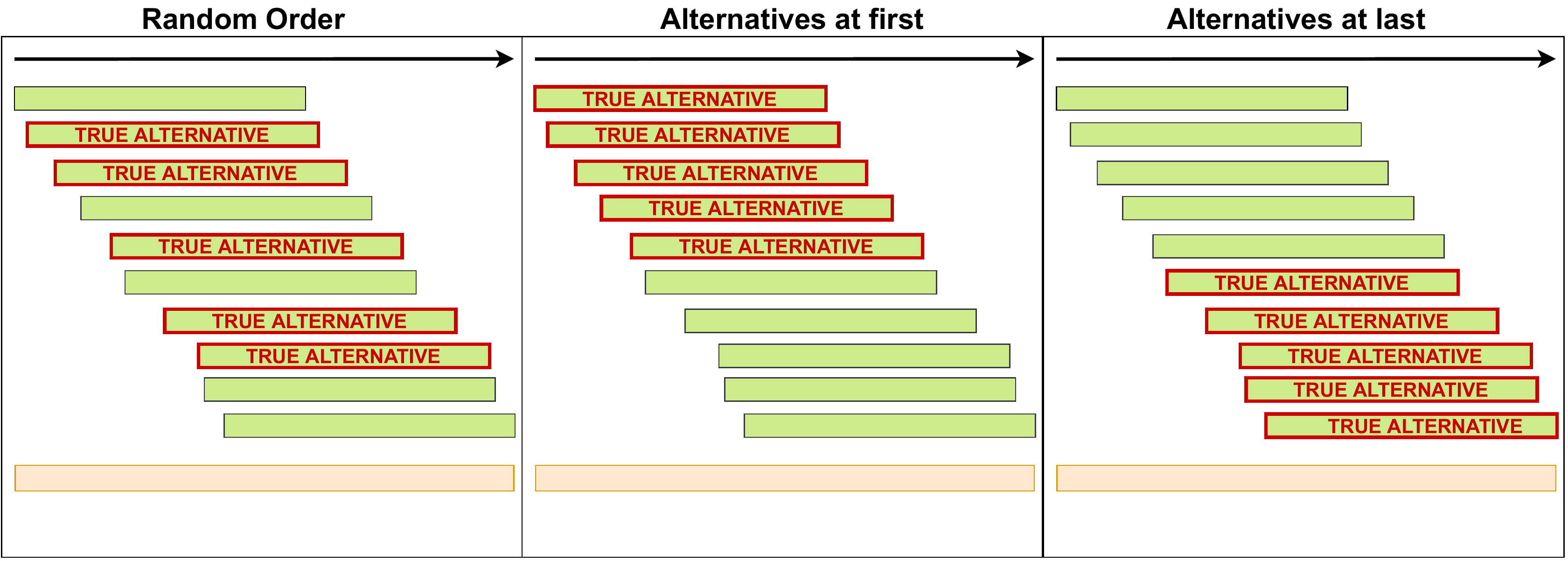}
\caption{Illustration of different orders of alternative hypotheses for the simulation studies.}
\label{fig:scenarios}
\end{figure}

In the simulations described in the manuscript we considered three different orders of alternative hypotheses (see Fig. \ref{fig:scenarios}): 
\begin{itemize}
\item Random order: In the simulations for each hypothesis the probability for a null hypothesis is $\pi_0$. Thus for an individual simulation step, the number of alternatives may differ.
\item Alternatives at first: The simulated platform trial starts with $m_1=K-\pi_0 K$ alternatives followed by $K-m_1$ true null hypotheses. 
\item Alternatives at last: The simulated platform trials starts with $K-m_1$ true null hypotheses followed by $m_1$ alternative hypotheses. 
\end{itemize}

\newpage

\section{Supplemental figures}
\subsection*{Comparison of LOND methods}

\begin{figure}[h!]
\centering
\includegraphics[width=15cm]{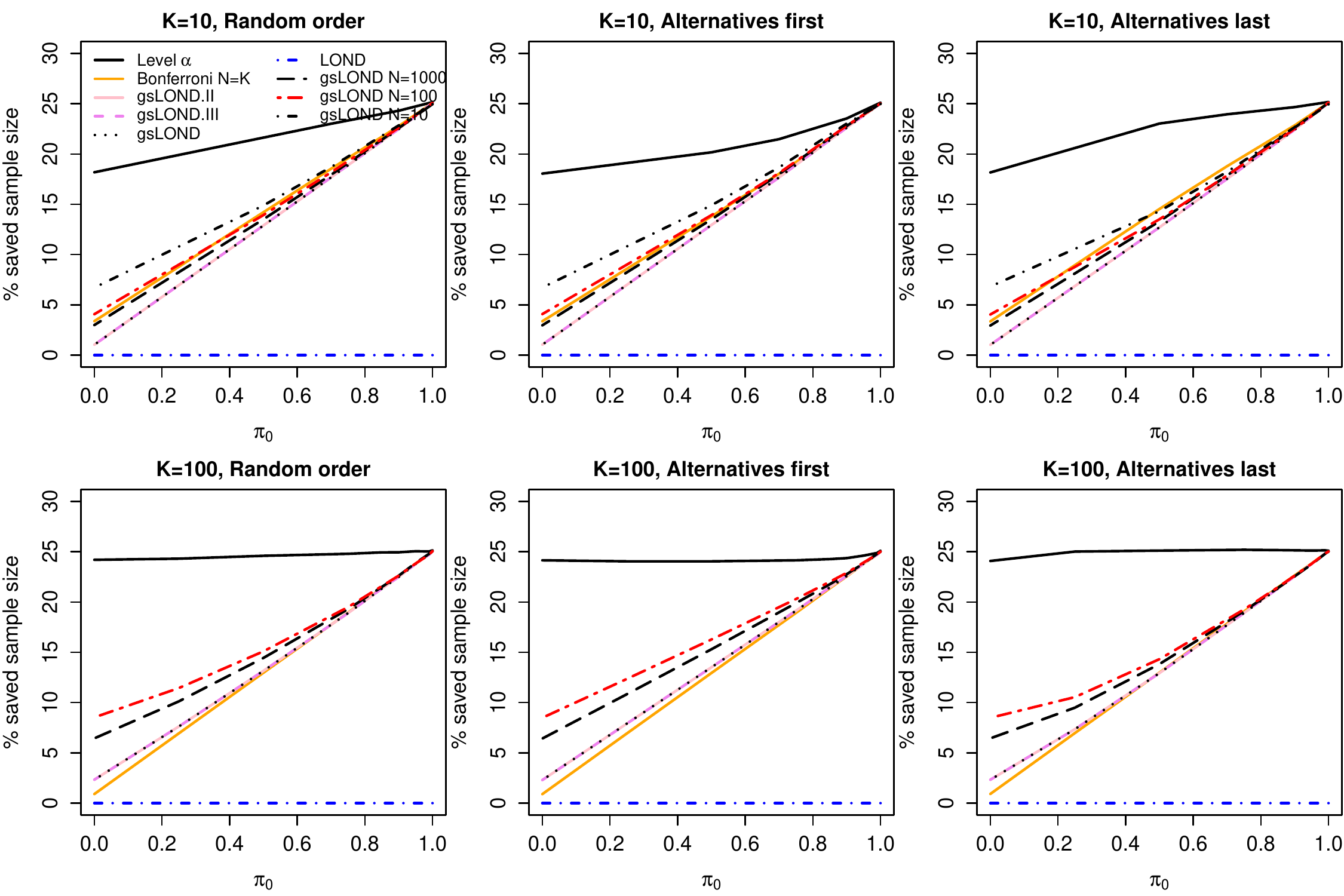}
\caption{\% saved sample size for CC scenario as a function of $\pi_0$ for the level-$\alpha$ and the Bonferroni procedure (with $N=K$) and the four LOND procedures LOND, gsLOND, gsLOND.II, and gsLOND.III. O'Brien Fleming design, $N=\{10,100,1000,\infty\}$, $\Delta=0.6$, $\alpha=0.025$, $\alpha^F=0.5$. The four LOND procedures can hardly be distinguished as the power values are very similar. Thus, for $N<\infty$, only the power for gsLOND is depicted.}
\label{fig:universe}
\end{figure}

\newpage
\begin{figure}[h!]
\centering
\includegraphics[width=15cm]{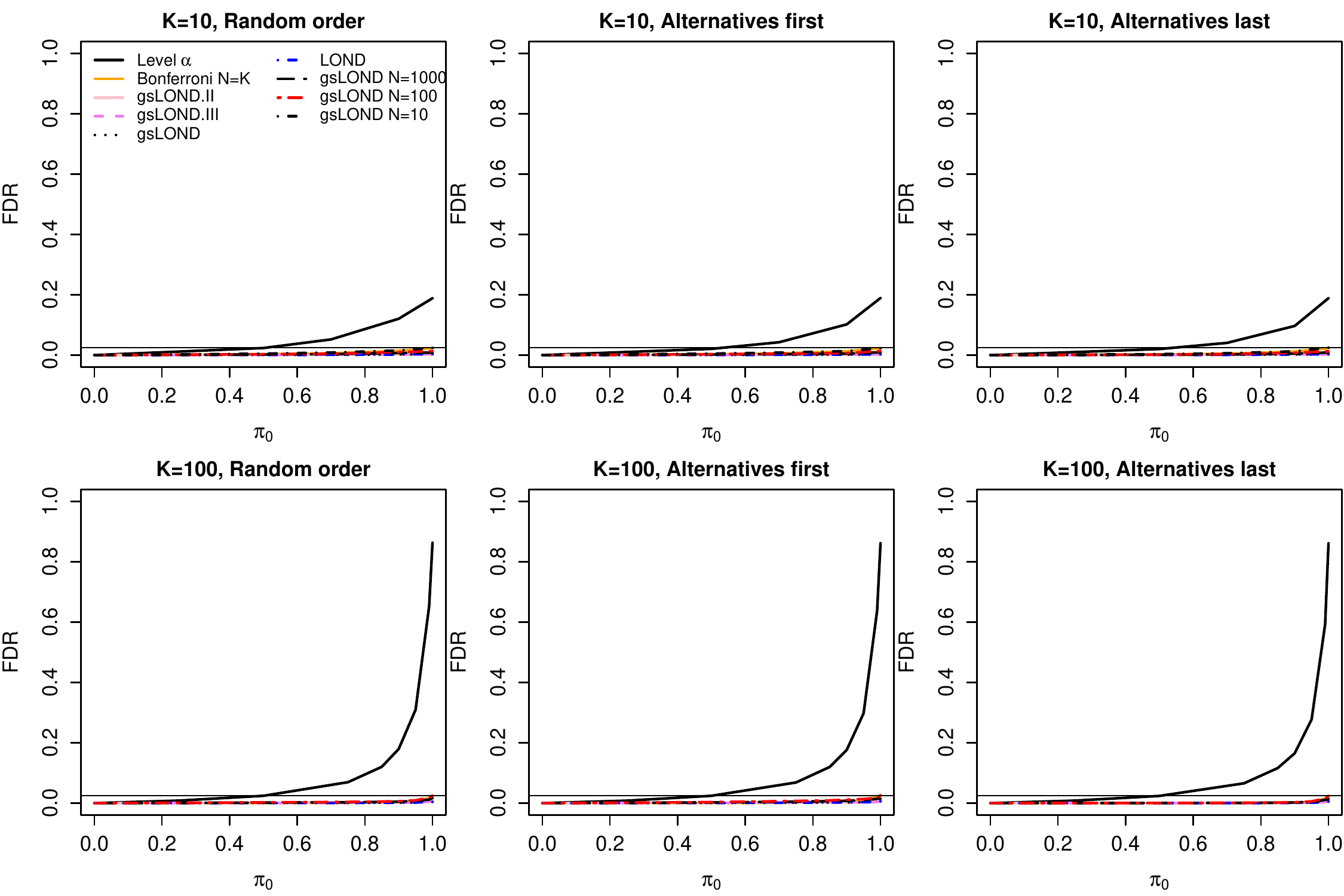}
\caption{Actual FDR for NCC+CC scenario as a function of $\pi_0$ for the level-$\alpha$ and the Bonferroni procedure (with $N=K$) and the four LOND procedures LOND, gsLOND, gsLOND.II, and gsLOND.III. O'Brien Fleming design, $N=\{10,100,1000,\infty\}$, $\Delta=0.6$, $\alpha=0.025$, $\alpha^F=0.5$. The four LOND procedures can hardly be distinguished as the power values are very similar. Thus, for $N<\infty$, only the power for gsLOND is depicted.}
\label{fig:universe}
\end{figure}

\newpage
\begin{figure}[h!]
\centering
\includegraphics[width=15cm]{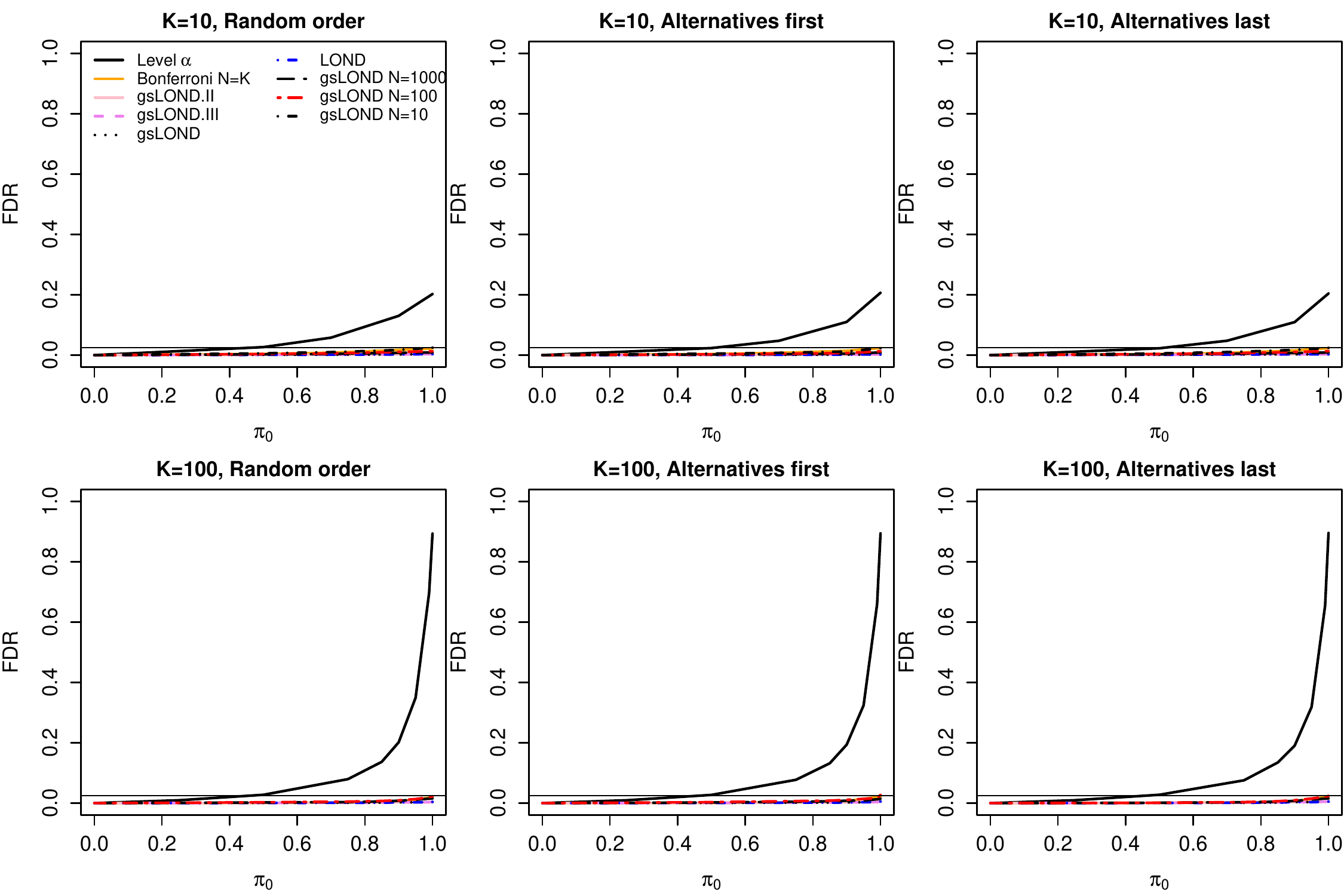}
\caption{Actual FDR for CC scenario as a function of $\pi_0$ for the level-$\alpha$ and the Bonferroni procedure (with $N=K$) and the four LOND procedures LOND, gsLOND, gsLOND.II, and gsLOND.III. O'Brien Fleming design, $N=\{10,100,1000,\infty\}$, $\Delta=0.6$, $\alpha=0.025$, $\alpha^F=0.5$. The four LOND procedures can hardly be distinguished as the power values are very similar. Thus, for $N<\infty$, only the power for gsLOND is depicted.}
\label{fig:universe}
\end{figure}

\newpage
\subsection*{Direct comparison of concurrent and non-concurrent controls}

\begin{figure}[h!]
\centering
\includegraphics[width=15cm]{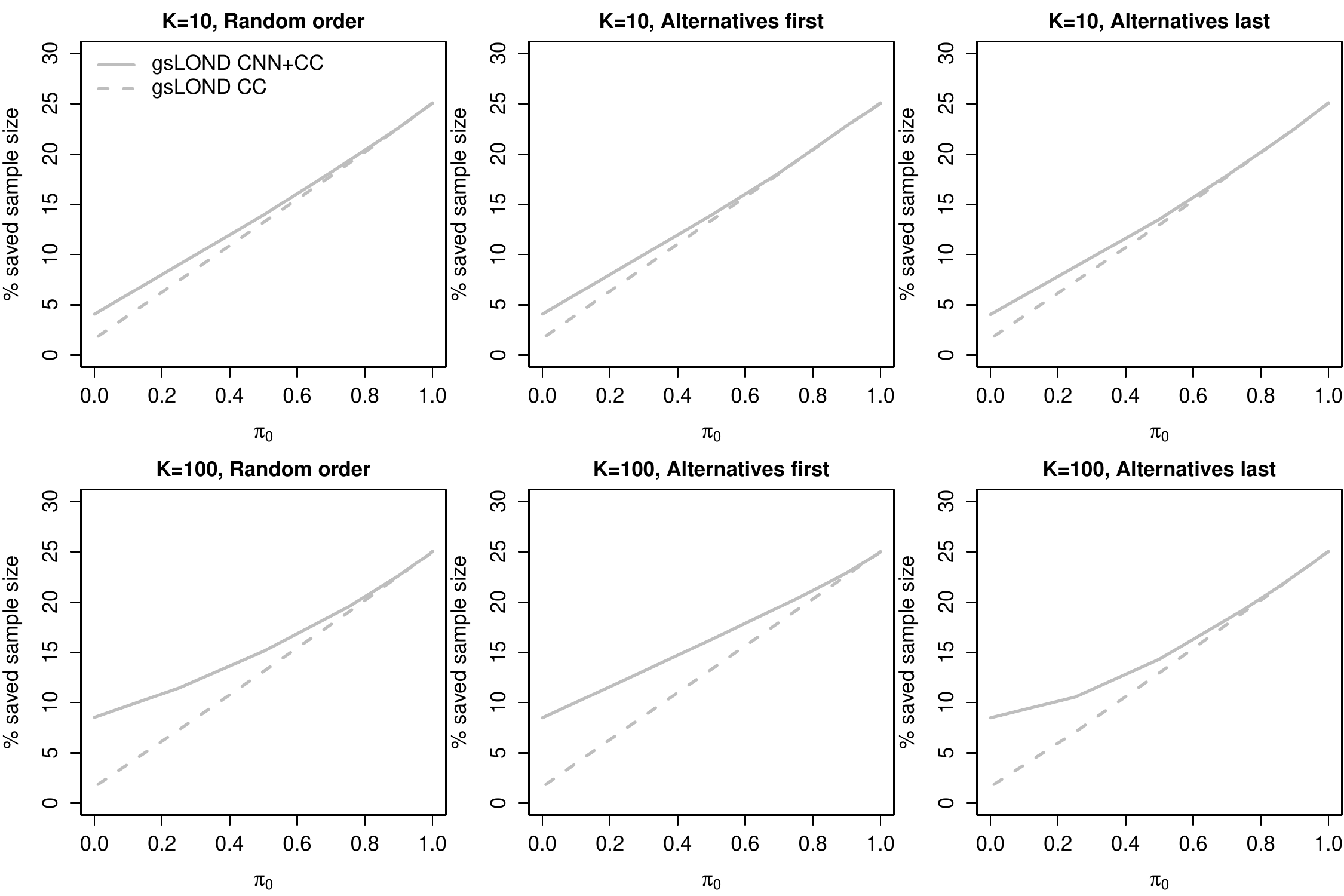}
\caption{Comparison of \% saved sample size for concurrent (CC) versus all (NCC+CC) controls as a function of $\pi_0$ for LOND and gsLOND (results for gsLOND.II and gsLOND.III are not depicted due to nearly identical power values); O'Brien Fleming design, $N=100$, $\Delta=0.6$, $\alpha=0.025$, $\alpha^F=0.5$.}
\label{fig:universe}
\end{figure}

\begin{figure}[h!]
\centering
\includegraphics[width=15cm]{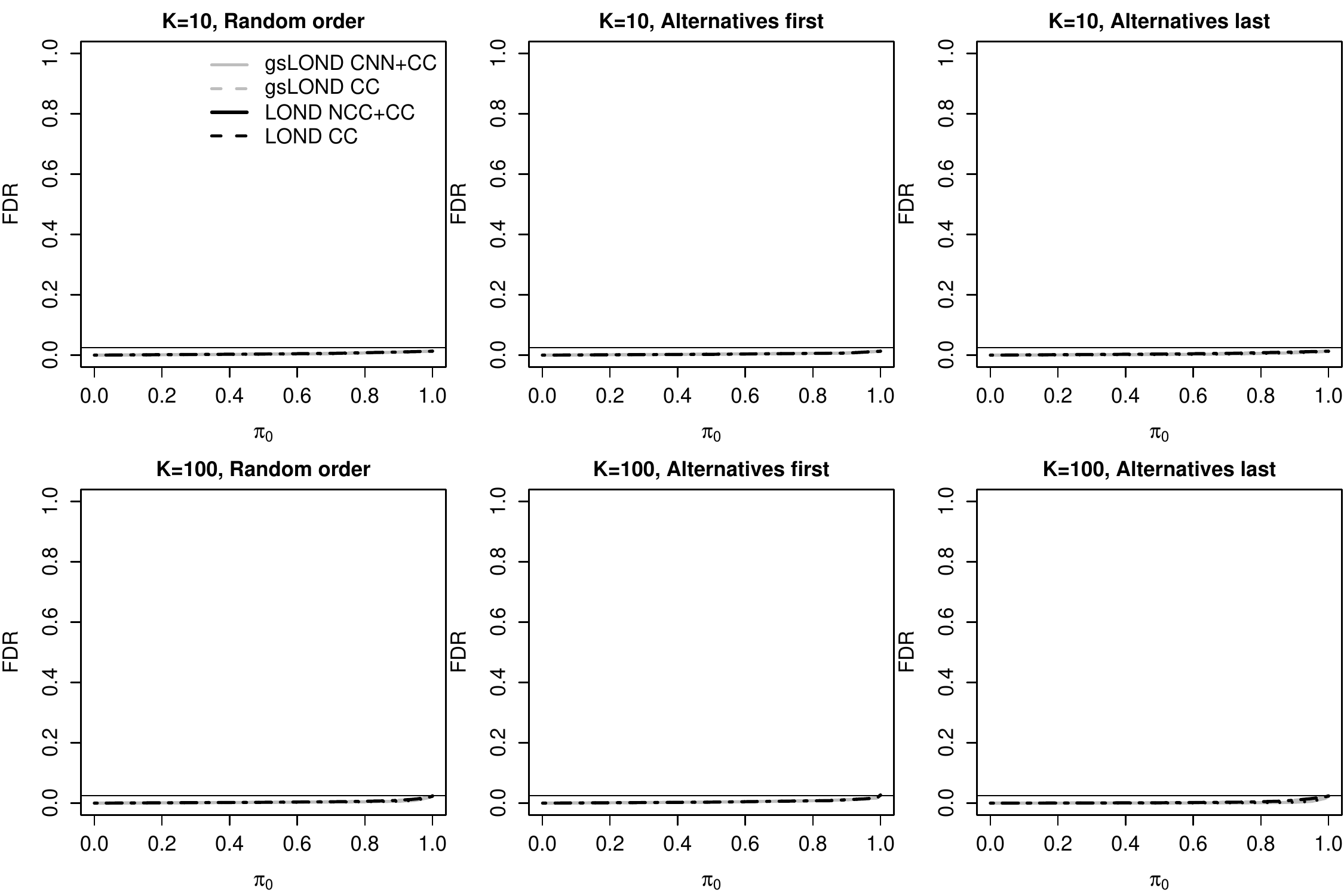}
\caption{Actual FDR for concurrent (CC) or all (NCC+CC) controls as a function of $\pi_0$ for gsLOND.II, and gsLOND.III; O'Brien Fleming design, $N=100$, $\Delta=0.6$, $\alpha=0.025$, $\alpha^F=0.5$.}
\label{fig:universe}
\end{figure}

\newpage
\subsection*{Distribution of significance level}

\begin{figure}[h!]
\centering
\includegraphics[width=15cm]{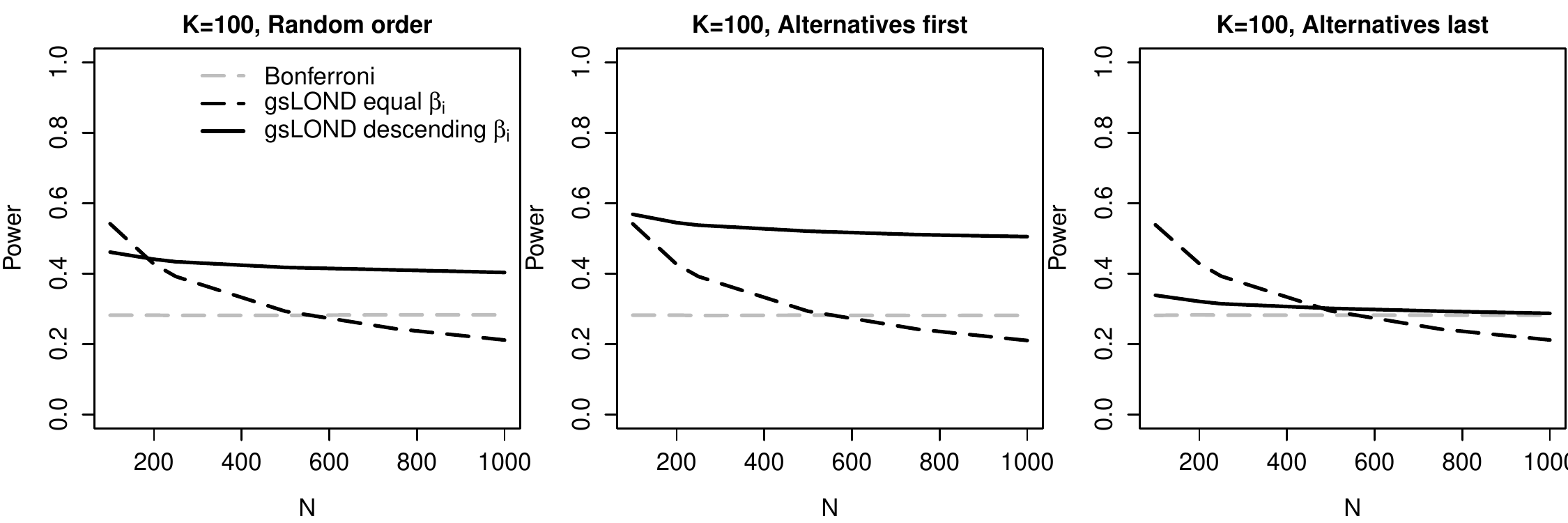}
\caption{Power comparison of gsLOND (CC controls) for two distributions of significance level as a function of the upper bound $N$ and Bonferroni with $N=K$. The values of $\beta$ are either derived by a descending or an uniform distribution. OBF design, $\Delta=0.6$, $\pi_0=0.5$, $K=100$, $\alpha=0.025$, $\alpha^F=0.5$.}
\label{fig:universe}
\end{figure}

\newpage
\subsection*{OBF versus PO design}

\begin{figure}[h!]
\centering
\includegraphics[width=10cm]{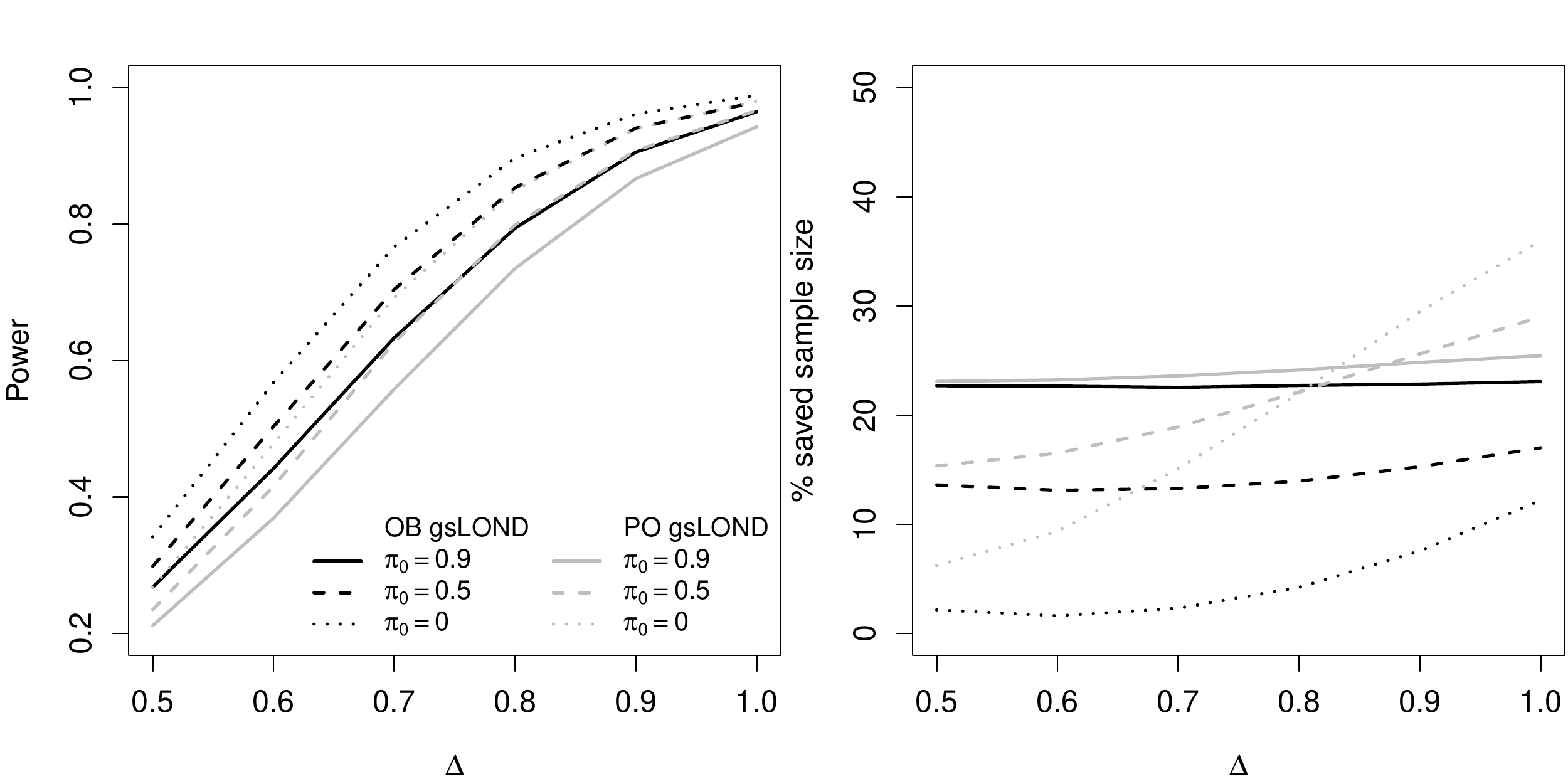}
\caption{Power values and \% saved sample size for OBF and PO designs of gsLOND as a function of the effect size for $\pi_0=\{0,0.5,0.9\}$ and $K=10$, $N=$, $\alpha=0.025$, $\alpha^F=0.5$, CC controls.}
\label{fig:universe}
\end{figure}

\begin{figure}[h!]
\centering
\includegraphics[width=10cm]{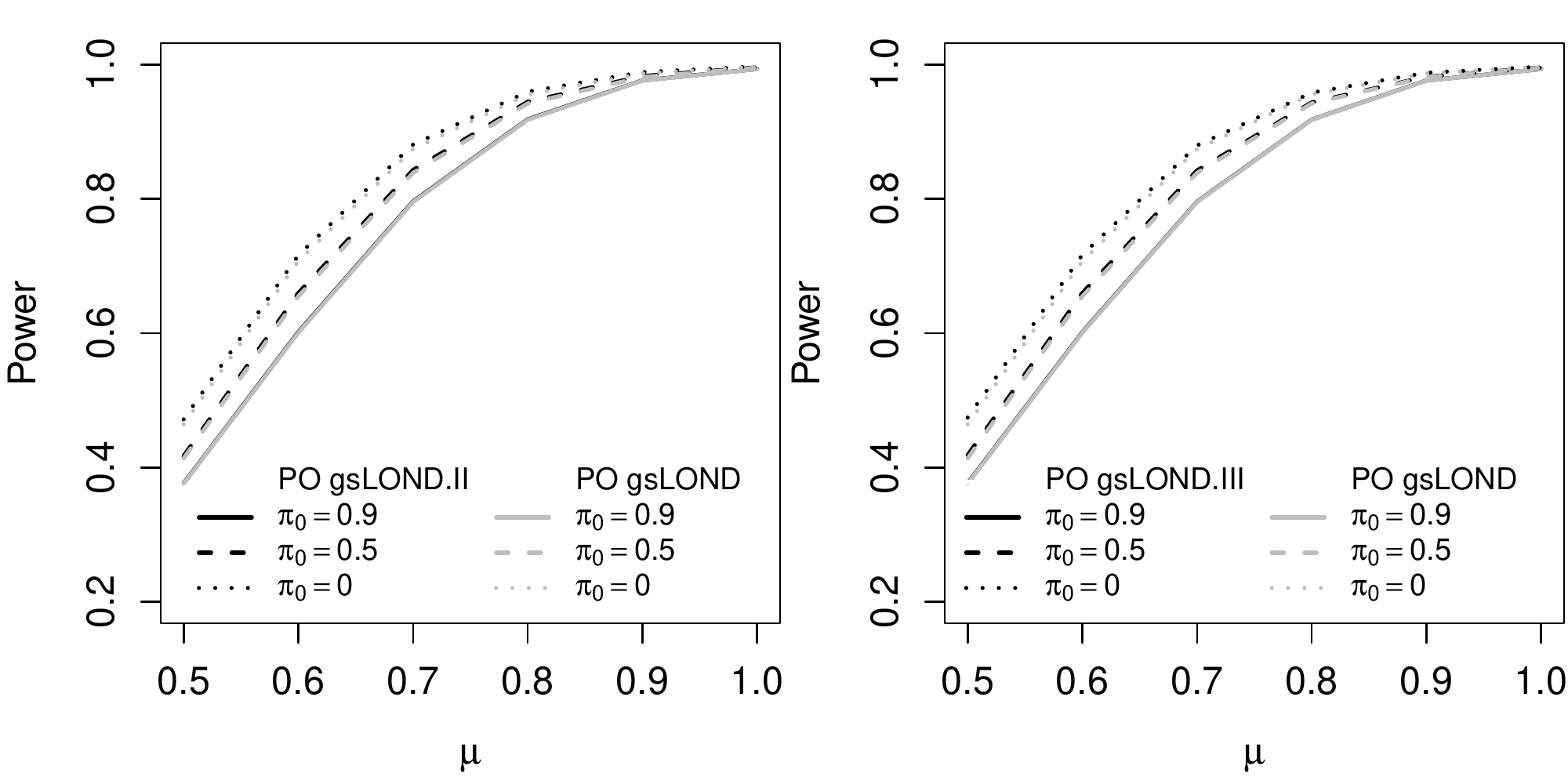}
\caption{Power values for PO designs of gsLOND and gsLOND.III as a function of the effect size for $\pi_0=\{0,0.5,0.9\}$ and $K=10$, $N=$, $\alpha=0.025$, $\alpha^F=0.5$.}
\label{fig:universe}
\end{figure}

\newpage
\subsection*{Inclusion of additional treatments for stopping in the interim analysis}

\begin{figure}[h!]
\centering
\includegraphics[width=10cm]{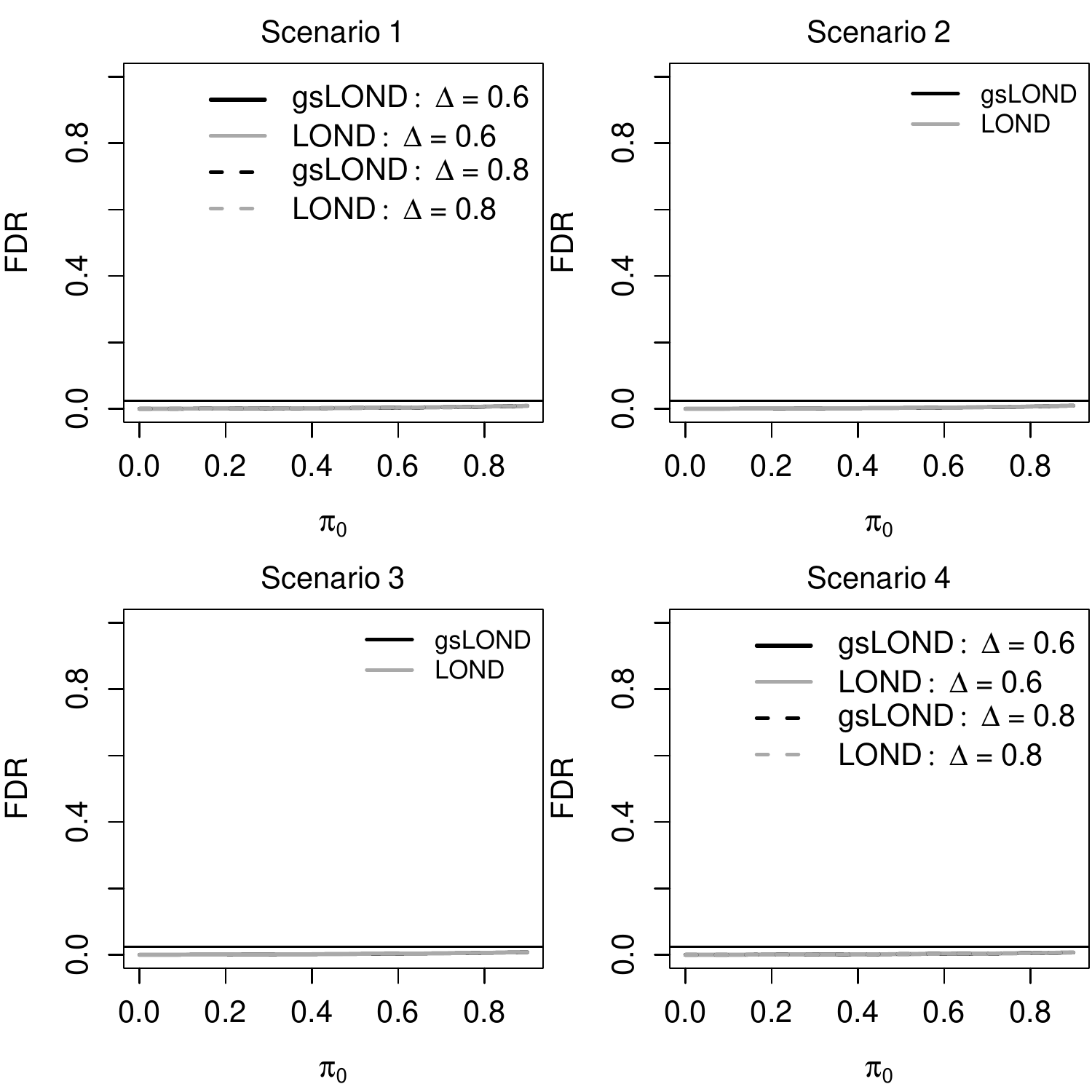}
\caption{Actual FDR with a fixed budget of the platform trial (NCC+CC) as a function of $\pi_0$ for the LOND and gsLOND. If a treatment arm is stopped early, an additional treatment is included. Scenario 1: ratio of true and false alternatives and effect size remains constant, $\Delta=0.6$ and 0.8. Scenario 2: distributed effect sizes of $\Delta=\{0.4, 0.8, 1.2\}$. Scenario 3: $\pi_0$ decreases by 1/80 for each new treatment. Scenario 4: $\Delta$ increases to $\Delta=1$ for additional alternative hypotheses. Initial $K=10$, $N=100$, $\alpha=0.025$, $\alpha^F=0.5$, OBF design. }
\label{fig:universe}
\end{figure}

\begin{figure}[h!]
\centering
\includegraphics[width=10cm]{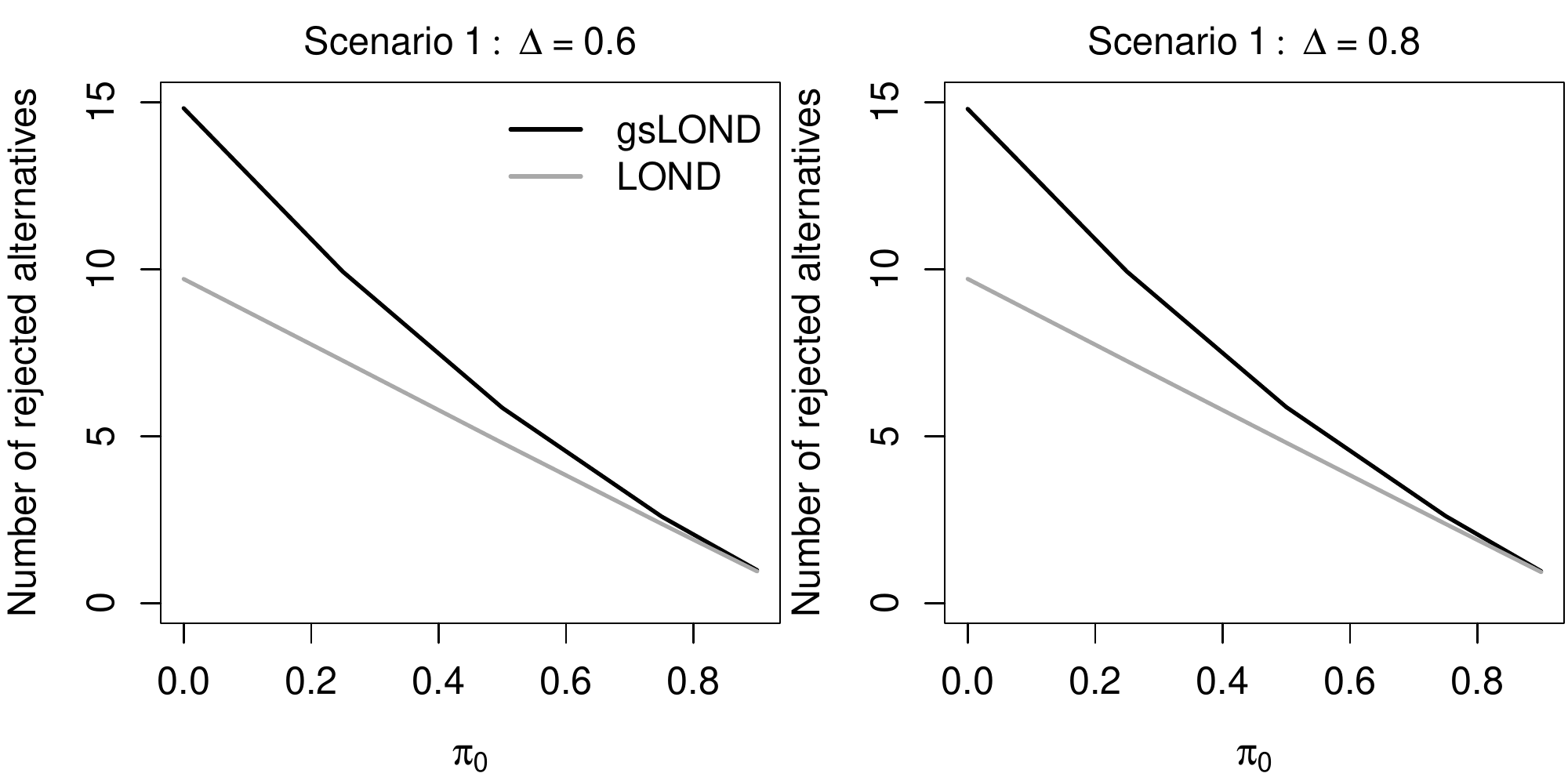}
\caption{Comparison of number of rejected alternatives with a fixed budget of the platform trial (NCC+CC) as a function of $\pi_0$ for the LOND and gsLOND. If a treatment arm is stopped early, an additional treatment is included. Scenario 1: ratio of true and false alternatives and effect size remains constant, $\Delta=0.6$ and 0.8. Initial $K=10$, $N=100$, $\alpha=0.025$, $\alpha^F=0.5$, OBF design. }
\label{fig:universe}
\end{figure}

\begin{figure}[h!]
\centering
\includegraphics[width=10cm]{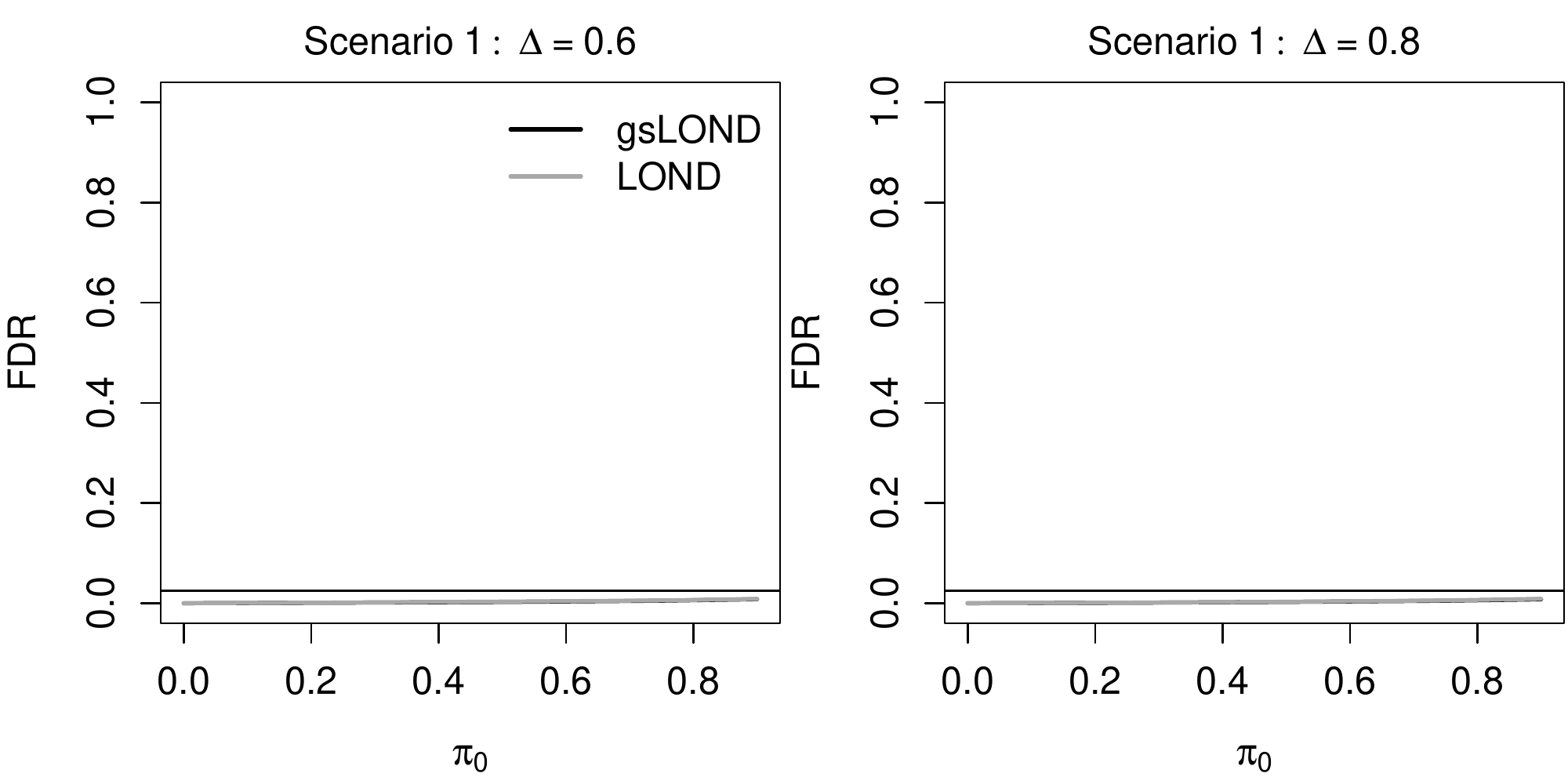}
\caption{Actual FDR with a fixed budget of the platform trial (NCC+CC) as a function of $\pi_0$ for the LOND and gsLOND. If a treatment arm is stopped early, an additional treatment is included. Scenario 1: ratio of true and false alternatives and effect size remains constant, $\Delta=0.6$ and 0.8. Initial $K=10$, $N=100$, $\alpha=0.025$, $\alpha^F=0.5$, OBF design. }
\label{fig:universe}
\end{figure}